\documentclass[12pt]{article}

\usepackage{mathrsfs,psfrag}
\usepackage{amsfonts}
\usepackage{graphicx}
\usepackage{mathrsfs}
\usepackage{amsmath}
\usepackage{algorithm}
\usepackage{algpseudocode}
\usepackage{amssymb}
\usepackage{float}
\usepackage{subfig}
\usepackage{rotating,color}
\usepackage{xcolor,natbib,epstopdf}
\usepackage{appendix}
	\renewcommand\appendix{\par
	\setcounter{section}{0}
	\setcounter{subsection}{0}
	\setcounter{lemma}{0}
	\setcounter{equation}{0}
	\renewcommand{\thelemma}{\arabic{lemma}}
	\gdef\thesection{Appendix \Alph{section}}
	\gdef\thesubsection{\Alph{section}.\arabic{subsection}}}

\newtheorem{theorem}{ \noindent T{\footnotesize HEOREM}}[section]

\newtheorem{lemma}{ \noindent L{\footnotesize EMMA}}[section]
\newtheorem{coro}{ \noindent C{\footnotesize OROLLARY}}[section]

\def\bm {\boldsymbol}
\def\u{\bm u}
\def\v{\bm v}
\def\bms{{\bf \Sigma}}
\def\S{{\bf S}}
\def\K{{\bf K}}
\def\I{{\bf I}}
\def\U{\bm U}
\def\bth{\bm \theta}
\def\hbth{\hat{\bth}}
\def\bmu{\bm \mu}
\def\X{\bm X}
\def\Y{\bm Y}
\def\var{{\rm Var}}
\def\cov{{\rm Cov}}
\def\tr{{\rm tr}}

\def\cd{\mathop{\rightarrow}\limits^{d}}

\title
{\bf Spatial Sign based Principal Component Analysis for High Dimensional Data}
\author{Ping Zhao$^1$, Hongfei Wang$^2$ and Long Feng$^3$\\
	$^1$ School of Mathematical Sciences, Tianjin Polytechnic University\\
	$^2$ School of Statistics and Data Science, Nanjing Audit university\\
	$^3$ School of Statistics and Data Science, KLMDASR, LEBPS, and LPMC,\\ Nankai University\\
}

\date{}

\begin{document}
\maketitle
\def\spacingset#1{\renewcommand{\baselinestretch}%
{#1}\small\normalsize} \spacingset{1}

\begin{abstract}
This article focuses on the robust principal component analysis (PCA) of high-dimensional data with elliptical distributions. We investigate the PCA of the sample spatial-sign covariance matrix in both nonsparse and sparse contexts, referring to them as SPCA and SSPCA, respectively. We present both nonasymptotic and asymptotic analyses to quantify the theoretical performance of SPCA and SSPCA. In sparse settings, we demonstrate that SSPCA, implemented through a combinatoric program, achieves the optimal rate of convergence. Our proposed SSPCA method is computationally efficient and exhibits robustness against heavy-tailed distributions compared to existing methods. Simulation studies and real-world data applications further validate the superiority of our approach.

{\it Keywords}:  Elliptical distribution; High dimensional data; Sparse principal component analysis; Spatial-sign covariance matrix.
\end{abstract}

\newpage
\spacingset{1.9} 

\section{Introduction}
PCA (Principal Component Analysis) is a widely used statistical method for data dimensionality reduction. It transforms high-dimensional data into a lower-dimensional space while preserving as much of the original data's variability as possible. PCA achieves this by identifying the directions of maximum variance in the data, known as principal components, and projecting the data onto these directions. This process removes redundant information and noise, making the data easier to handle and visualize. PCA is commonly applied in fields such as machine learning { \citep{balcan2016communication}}, image processing { \citep{chan2015pcanet}}, and finance { \citep{lan2019a,yang2025}}, where dealing with high-dimensional data is common.

The classical PCA method can encounter significant difficulties when the number of input variables $ p $ is not substantially smaller than the sample size $ n $. Specifically, PCA becomes inconsistent when the ratio $ p/n $ converges to some $ \gamma $ within the interval (0, 1), as noted by \cite{Johnstone2009}. Furthermore, PCA's performance deteriorates even more dramatically when $ p $ is significantly larger than $ n $. To address this challenge, we invoke the assumption of sparsity. One form of sparsity assumption pertains to the spectrum of the covariance matrix, as explored in works such as \cite{Johnstone2009}, \cite{Baik2006}, \cite{Paul2007}, \cite{Nadler2008}, \cite{Birnbaum2013}, and \cite{Cai2013}. Another, more widely adopted assumption, focuses on the sparsity of the eigenvectors of the covariance matrix. Sparsity in the loadings offers a distinct advantage in interpretability, as it implies that each principal component is influenced by a limited subset of input variables. In this study, we focus on this latter assumption.

In the context of sparse settings, numerous sparse PCA methods have been explored in the literature. \cite{Jolliffe2003modified} and \cite{Zou2006}  approached principal component analysis as a regression-type optimization problem and incorporated lasso-type penalties for parameter estimation. \cite{Shen2008} and \cite{Witten2009} leveraged the relationship between PCA and singular value decomposition (SVD) to extract sparse loadings through iterative thresholding. \cite{journee2010generalized} introduced Gpower, a generalized power method for sparse PCA, by reformulating PCA with sparsity-inducing penalties as the maximization of a convex function over a sphere. \cite{ZhangElGhaoui2011} proposed a greedy search algorithm for finding principal submatrices of the covariance matrix. \cite{VuChoLeiRohe2013} formulated the sparse principal subspace problem as a semidefinite program with a Fantope constraint and developed the Fantope Projection and Selection (FPS) algorithm to solve it. \cite{ma2013sparse} and \cite{YuanZhang2013} suggested modified versions of the power method for estimating eigenvectors and principal subspaces. For a thorough overview of sparse PCA, readers are referred to \cite{Zou2018}.

One limitation of the aforementioned PCA and sparse PCA methods is their reliance on the assumption of Gaussian or sub-Gaussian distributions. When observations exhibit heavy-tailed behavior, these estimators may not be consistent. To tackle this problem, numerous studies have proposed replacing the sample covariance matrix with a robust covariance matrix. Examples include the works of \cite{hubert2005robpca}, \cite{Croux2013}, \cite{Han2014}, \cite{hubert2016sparse}, \cite{han2018eca}. These approaches aim to enhance the robustness of PCA and sparse PCA methods by utilizing robust covariance matrices that are less sensitive to outliers and heavy-tailed distributions. Specially, \cite{Han2014} employed the marginal Kendall's tau statistic to estimate the correlation matrix under the semiparametric transelliptical family. However, as highlighted in \cite{han2018eca}, this method has two primary drawbacks: it only estimates the correlation matrix rather than the covariance matrix, and the sign sub-Gaussian condition is not straightforward to verify. To overcome these limitations, \cite{han2018eca} proposed using the multivariate Kendall's tau matrix as a substitute for the sample covariance matrix to estimate eigenvectors under the elliptical model and various settings.

Under the assumption of an elliptical distribution, \cite{marden1999some} demonstrated that both the population multivariate Kendall's tau matrix and the spatial-sign covariance matrix share the same eigenspace as the covariance matrix. Consequently, these two matrices have been widely used in the literature to estimate principal components in low-dimensional settings. Notable contributions include the works of \cite{locantore1999robust}, \cite{marden1999some}, \cite{visuri2000sign}, \cite{Croux2002}, \cite{Taskinen2012}. It is important to note that the population multivariate Kendall's tau matrix is equivalent to the population spatial-sign covariance matrix. However, when considering their sample counterparts, the computational complexity of the sample multivariate Kendall's tau matrix ($n^2d^2$) is significantly higher than that of the sample spatial-sign covariance matrix ($nd^2$), particularly for large sample sizes. Therefore, it is of great interest to analyze the performance of the principal component estimator using the sample spatial-sign covariance matrix in high-dimensional scenarios.

For elliptical distributions, classic spatial-sign-based procedures have proven to be highly robust and efficient in traditional multivariate analysis, as overviewed by \citet{Oja2010Multivariate}. Recent literature has also shown that these spatial-sign-based procedures excel in high-dimensional settings. Specifically, \citet{wang2015high}, \citet{feng2016}, and \citet{f2021} have proposed spatial-sign-based test procedures for the high-dimensional one-sample location problem. Additionally, \citet{Feng2016Multivariate} and \citet{h2022} have addressed the high-dimensional two-sample location problem using spatial-sign-based methods. Moreover, \citet{zou2014multivariate}, \citet{fl2017} and \cite{zhang2022robust} extended the spatial-sign-based method to the high-dimensional sphericity test, while \citet{paindaveine2016high} and \cite{zhao2023spatial} considered high-dimensional white noise tests.

In this paper, we investigate principal component analysis using the spatial-sign covariance matrix in high-dimensional settings. Firstly, we establish theoretical results for Spatial-sign based Principal Component Analysis (SPCA) in the nonsparse scenario. We demonstrate that the rate of convergence of the eigenvector comprises two components: one is comparable to that of the Elliptical Component Analysis (ECA) proposed by \citep{han2018eca}, i.e., $O_p(\sqrt{r^*(\bms) \log d/n})$, and the other is influenced by the consistency of the spatial median. This is not unexpected, as the sample spatial-sign covariance matrix requires the estimation of the location parameter, whereas the sample multivariate Kendall's tau matrix does not. Fortunately, under certain mild conditions, we can show that the second component is of a smaller order compared to the first. Secondly, in the sparse setting, we propose a Sparse Spatial-sign based Principal Component Analysis (SSPCA) through a combinatorial program and demonstrate that it can achieve the minimax optimal rate of convergence. Thirdly, we present a computationally efficient algorithm based on the truncated power method proposed by \citep{YuanZhang2013}. We also consider two initial estimators. One is the simple eigenvectors of the sample spatial-sign covariance matrix, which is very easily computed. The other is using the Fantope projectoin \citep{VuChoLeiRohe2013}. Lastly, we also provide a procedure for estimating the tuning parameter that controls the sparsity level.

Simulation studies indicate that our proposed methods exhibit robustness in handling heavy-tailed distributions. When compared to ECA, our SSPCA not only computes more rapidly but also demonstrates greater efficiency. These findings align with those reported in \cite{feng2018power}, which suggest that rank-based methods are less efficient than sign-based methods in high-dimensional contexts. Furthermore, our newly proposed method for determining the number of nonzero components in eigenvectors remains consistent as sample sizes increase. Applications to real data further underscore the advantages of our approach.

The remainder of this article is structured as follows. Section 2 introduces SPCA in non-sparse settings, while Section 3 presents the SSPCA method in sparse settings. Simulation studies are discussed in Section 4, and real data applications are examined in Section 5. Section 6 concludes the article, and all the detailed proofs are provided in the Appendix.

{\it Notation}:
Here we use the same notations as \cite{han2018eca}. Let $\mathbf{M}=\left[\mathbf{M}_{j k}\right] \in \mathbb{R}^{d \times d}$ be a symmetric matrix and $\boldsymbol{v}=\left(v_1, \ldots, v_d\right)^T \in \mathbb{R}^d$ be a vector. We denote $\boldsymbol{v}_I$ to be the subvector of $\boldsymbol{v}$ whose entries are indexed by a set $I$, and $\mathbf{M}_{I, J}$ to be the submatrix of $\mathbf{M}$ whose rows are indexed by $I$ and columns are indexed by $J$. We denote $\operatorname{supp}(v):=\left\{j: v_j \neq 0\right\}$. For $0<q<\infty$, we define the $\ell_q$ and $\ell_{\infty}$ vector norms as $\|\boldsymbol{v}\|_q:=\left(\sum_{i=1}^d\left|v_i\right|^q\right)^{1 / q}$ and $\|\boldsymbol{v}\|_{\infty}:=\max _{1 \leq i \leq d}\left|v_i\right|$. We denote $\|\boldsymbol{v}\|_0:=\operatorname{card}(\operatorname{supp}(\boldsymbol{v}))$. We define the matrix entry-wise maximum value and Frobenius norms as $\|\mathbf{M}\|_{\text {max }}:=\max \left\{\left|\mathbf{M}_{i j}\right|\right\}$ and $\|\mathbf{M}\|_{\mathrm{F}}=\left(\sum \mathbf{M}_{j k}^2\right)^{1 / 2}$. Let $\lambda_j(\mathbf{M})$ be the $j$ th largest eigenvalue of $\mathbf{M}$. If there are ties, $\lambda_j(\mathbf{M})$ is any one of the eigenvalues such that any eigenvalue larger than it has rank smaller than $j$, and any eigenvalue smaller than it has rank larger than $j$. Let $\boldsymbol{u}_j(\mathbf{M})$ be any unit vector $\boldsymbol{v}$ such that $\boldsymbol{v}^T \mathbf{M} \boldsymbol{v}=\lambda_j(\mathbf{M})$. Without loss of generality, we assume that the first nonzero entry of $\boldsymbol{u}_j(\mathbf{M})$ is positive. We denote $\|\mathbf{M}\|_2$ to be the spectral norm of $\mathbf{M}$ and $\mathbb{S}^{d-1}:=\left\{\boldsymbol{v} \in \mathbb{R}^d:\|\boldsymbol{v}\|_2=1\right\}$ to be the $d$-dimensional unit sphere. We define the restricted spectral norm $\|\mathbf{M}\|_{2, s}:=\sup _{\boldsymbol{v} \in \mathbb{S}^{d-1},\|\boldsymbol{v}\|_0 \leq s}\left|\boldsymbol{v}^T \mathbf{M} \boldsymbol{v}\right|$, so for $s=d$, we have $\|\mathbf{M}\|_{2, s}=\|\mathbf{M}\|_2$. We denote $f(\mathbf{M})$ to be the matrix with entries $[f(\mathbf{M})]_{j k}=f\left(\mathbf{M}_{j k}\right)$. We denote $\operatorname{diag}(\mathbf{M})$ to be the diagonal matrix with the same diagonal entries as $\mathbf{M}$. Let $\mathbf{I}_d$ represent the $d$ by $d$ identity matrix. For any two numbers $a, b \in \mathbb{R}$, we denote $a \wedge b:=\min \{a, b\}$ and $a \vee b:=\max \{a, b\}$. For any two sequences of positive numbers $\left\{a_n\right\}$ and $\left\{b_n\right\}$, we write $a_n \asymp b_n$ if $a_n=O\left(b_n\right)$ and $b_n=O\left(a_n\right)$. We write $b_n=\Omega\left(a_n\right)$ if $a_n=O\left(b_n\right)$, and $b_n=\Omega^o\left(a_n\right)$ if $b_n=\Omega\left(a_n\right)$ and $b_n \not\asymp a_n$. For any random variable $X \in \mathbb{R}$, we define the sub-Gaussian $\left(\|\cdot\|_{\psi_2}\right.$ ) and sub-exponential norms $\left(\|\cdot\|_{\psi_1}\right.$ ) of $X$ as follows:
$
 \|X\|_{\psi_2}:=\sup _{k \geq 1} k^{-1 / 2}\left(E|X|^k\right)^{1 / k} \text { and } 
 \|X\|_{\psi_1}:=\sup _{k \geq 1} k^{-1}\left(E|X|^k\right)^{1 / k}.
$
Any $d$-dimensional random vector $X \in \mathbb{R}^d$ is said to be sub-Gaussian distributed with the sub-Gaussian constant $\sigma$ if
$
\left\|\boldsymbol{v}^T \boldsymbol{X}\right\|_{\psi_2} \leq \sigma, \text { for any } \boldsymbol{v} \in \mathbb{S}^{d-1}.
$
For any two vectors $\boldsymbol{v}_1, \boldsymbol{v}_2 \in  \mathbb{S}^{d-1}$, let $\sin \angle\left(\boldsymbol{v}_1, \boldsymbol{v}_2\right)$ be the sine of the angle between $\boldsymbol{v}_1$ and $\boldsymbol{v}_2$, with
$
\left|\sin \angle\left(\boldsymbol{v}_1, \boldsymbol{v}_2\right)\right|:=\sqrt{1-\left(\boldsymbol{v}_1^T \boldsymbol{v}_2\right)^2}.
$

\section{SPCA: Nonsparse Setting}

Suppose $d$-dimensional random vector $\boldsymbol{X}$ follows elliptical distribution $EC_d(\boldsymbol{\mu},\boldsymbol{\Sigma},\xi)$, i.e.
$$
\boldsymbol{X} \stackrel{\mathrm{d}}{=} \boldsymbol{\mu}+\xi \mathbf{A} \boldsymbol{U},
$$
where $\boldsymbol{\mu} \in \mathbb{R}^d$, $\boldsymbol{U}$ is a uniform random vector on the unit sphere in $\mathbb{R}^q, \xi \geq 0$ is a scalar random variable independent of $\boldsymbol{U}$, and $\mathbf{A} \in \mathbb{R}^{d \times q}$ is a deterministic matrix satisfying $\mathbf{A A}^T=\boldsymbol{\Sigma}$ and $\boldsymbol{\Sigma} \in \mathbb{R}^{d \times d}$ with rank $(\boldsymbol{\Sigma})=q \leq d$.  Here, $\boldsymbol{\Sigma}$ is called the scatter matrix. In this article, we only consider continuous elliptical distributions with $P(\xi=0)=0$. Similar to \cite{han2018eca}, we also assume $E(\xi^2)=q<\infty$ so that $\cov(\X)=\bms$. In fact, our proposed methods still work even when $E(\xi^2)=\infty$.

The Spatial-Sign Covariance Matrix is defined as
\begin{align}
{\bf S}\doteq E\left(U(\boldsymbol{X}_i-\bm \mu)U(\boldsymbol{X}_i-\bm \mu)^T\right)
\end{align}
where $U(\bm x)=\bm x/\|\bm x\|{ _2}I(\bm x\not=\bm 0)$ is the spatial sign function.
By Theorem 4.4 in \cite{Oja2010Multivariate}, we know that the eigenspace of the spatial sign covariance matrix ${\bf S}$ is identical to the eigenspace of the covariance matrix ${\bf \Sigma}$. Note that, according to Lemma B.1 in \cite{han2018eca}, we know that the multivariate Kendall's tau matrix. i.e.
\begin{align}
{\bf K}=E\left(U(\bm X_i-\bm X_j)U(\bm X_i-\bm X_j)^T\right)
\end{align}
is equal to the spatial sign covariance matrix $\S$. So by Proposition 2.1 in \cite{han2018eca}, the eigenvalues of ${\bf S}$ is
\begin{align}
\lambda_j(\mathbf{S})=E\left(\frac{\lambda_j(\boldsymbol{\Sigma}) Y_j^2}{\lambda_1(\boldsymbol{\Sigma}) Y_1^2+\cdots+\lambda_q(\boldsymbol{\Sigma}) Y_q^2}\right)
\end{align}
if $rank({\bf S})=q$, where $\boldsymbol{Y}:=\left(Y_1, \ldots, Y_q\right)^T \sim N_q\left(\mathbf{0}, \mathbf{I}_q\right)$ is a standard multivariate Gaussian distribution. In addition, $\mathbf{S}$ and $\boldsymbol{\Sigma}$ share the same eigenspace with the same descending order of the eigenvalues. To estimate the spatial sign covariance matrix, we first need to estimate the location parameter $\bm \mu$. We often use the spatial median to estimate $\bm \mu$, i.e.
\begin{align}\label{emu}
\hat{\bm \mu}=\underset{\bm \mu \in \mathbb{R}^p}{\arg\min} \sum_{i=1}^n\|\bm X_i-\bm \mu\|_2.
\end{align}
Then the sample spatial sign covariance matrix is defined as
\begin{align}\label{hats}
\hat{{\bf S}}=\frac{1}{n}\sum_{i=1}^n U(\bm X_i-\hat{\bm \mu})U(\bm X_i-\hat{\bm \mu})^T
\end{align}
\cite{visuri2000sign} show that the influence functions of the sample spatial sign covariance matrix are uniformly bounded, indicating their robustness. In fact, the influence function of the sample spatial sign covariance matrix at a distribution $F$ symmetric around zero has the simple form
$
I F(\boldsymbol{x}, \hat \S, F)=U(\boldsymbol{x}) U(\boldsymbol{x})^T-\hat{\S}
$
and is seen to be constant in the radius $\|\mathbf{x}\|{ _2}$ of the contamination point $\boldsymbol{x}$.

In this section, we first do not assume the sparsity of $\u_1(\bms)$ and assume that $\lambda_1(\bms)$ is distinct. Consequently, we propose the leading eigenvector of $\hat{\S}$, ie. $\u_1(\hat \S)$ to estimate $\u_1(\S)=\u_1(\bms)$:

{\it The SPCA (Spatial-sign based Principal Component Analysis) estimator: $\u_1(\hat \S)$ (the leading eigenvector of $\S$).}

When the dimension $d$ is fixed and $rank(\bms)=q$, \cite{Croux2002} showed the asymptotic normality of ${\u}_1(\hat \S)$, i.e.
$
\sqrt{n}({\u}_1(\hat \S)-{ \u_1( \S)})\cd N(0,\sigma_{u1}^2)
$
where
$
\sigma_{u1}^2=\sum_{l \neq 1}\left[\frac{\lambda_l(\bms) \lambda_1(\bms) b_{1 l}}{\left(c_1  \lambda_1(\bms)-c_l  \lambda_l(\bms)\right)^2} \boldsymbol{u}_l \boldsymbol{u}_l^T\right]
$
where
$
c_l =E\left[u_{ l}^2 /\left(\gamma_1 u_1^2+\ldots+\gamma_q u_q^2\right)\right]
$
for $l=1, \ldots, q$, and for $1 \leq j, l \leq q$,
$
b_{1 l} =E\left[u_1^2 u_l^2 /\left(\gamma_1 u_1^2+\ldots+\gamma_q u_q^2\right)\right]^2
$
with $\left(u_1, \ldots, u_q\right)$ the components of a random variable $\boldsymbol{u}$, uniformly distributed on the periphery of a unit sphere, and $\gamma_1, \ldots, \gamma_p$ the standardized eigenvalues, that is $\gamma_j(\bms)=\lambda_j(\bms) /\left(\lambda_1(\bms)+\ldots, \lambda_q(\bms)\right)$.

To the best of our knowledge, as the dimension $d$ approaches infinity, there are currently no asymptotic results available for $\hat{\u}_1(\hat{\mathbf{S}})$. { To fill this gap,} we first analysis the convergence rate of $\u_1(\hat \S)$ in high dimensional settings. According to Davis-Kahan inequality \citep{Davis1970,Wedin1972}, to evaluate the convergence rate of $\u_1(\hat \S)$ to $\u_1(\S)$, we first study the convergence rate of $\hat{\S}$ to $\S$ under the spectral norm.

Define $\zeta_k=E(r_i^{-k})$, $r_i=\|\X_i-\bmu\|{ _2}$, $\nu_i=\zeta_1^{-1}r_i^{-1}$. We assume that
\begin{itemize}
\item[(A1)] $\zeta_k\zeta_1^{-k}< \zeta \in (0,\infty)$ for $k=1,2,3,4$ and all $d$.
\item[(A2)] $\lim\sup_d\|\S\|_2<1-\psi<1$ for some positive constant $\psi$.
\end{itemize}
Assumption (A1) is widely assumed in high dimensional spatial-sign based procedures, such as \cite{zou2014multivariate},\cite{Feng2016Multivariate}. By condition (A1), we have $E(\nu_i)=1, \sigma_\nu^2=\var(\nu_i)<\zeta-1, \kappa_\nu=E(\nu_i^4)<\zeta$. Assumption (A2) means that the maximum eigenvalue of $\S$ should be uniformly smaller than one, which is employed to guarantee the consistency of the spatial median. In the past decades, there are some literatures which established the consistency of the spatial median under different assumptions, such as \cite{zou2014multivariate}, \cite{cheng2019testing}, \cite{li2022asymptotic}. However, all the above papers need to assume the eigenvalues of $\bms$ are all bounded or $\tr(\bms^4)=o(\tr^2(\bms^2)$, which is too restrictive in principal component analysis. In contrast, Assumption (A2) is less stringent than these prior assumptions.

Let $r^*(\S)=\tr(\S)/\|\S\|_2=\frac{1}{\lambda_1(\S)}$, which is referred to as the effective rank of $\S$ in the literature \citep{vershynin2010, lounici2014high}. The next theorem establish the convergence rate of $\|\hat \S-\S\|_2$.

\begin{theorem}\label{th1}
Let $\boldsymbol{X}_1, \ldots, \boldsymbol{X}_n$ be $n$ independent observations of $\boldsymbol{X} \sim E C_d(\boldsymbol{\mu}, \boldsymbol{\Sigma}, \xi)$. Let $\widehat{\mathbf{S}}$ be the sample version of the spatial-sign covariance matrix defined in Equation (\ref{hats}). We have, for any $\alpha>0$, there exist a positive constant $C_S$, such that, for sufficient large $n$ and any $\delta\in (0,1)$,
\begin{align}\label{s2n}
\|\hat{\S}-\S\|_2\le \|\mathbf{S}\|_2 \sqrt{\frac{4\left(1+r^*(\mathbf{S})\right)(\log d+\log (1    / \alpha))}{n}}+C_S n^{-\frac{1}{2}(1+\delta)}
\end{align}
with probability larger than $1-\alpha$.
\end{theorem}

The initial term in (\ref{s2n}) bears a resemblance to the nonasymptotic bound for $\|\hat{\K}-\K\|{ _2}$ presented in Theorem 3.1 of \cite{han2018eca}. Here, $\hat{\K}$ represents the sample multivariate Kendall's tau estimator. Essentially, this term constitutes the nonasymptotic bound for $\|\tilde{\S}-\S\|{ _2}$, where $\tilde{\S} = \frac{1}{n}\sum_{i=1}^n U(\X_i-\bmu)U(\X_i-\bmu)^T$. This result is established using the matrix Bernstein inequality introduced by \cite{tropp2012}. The second term in (\ref{s2n}) arises from the convergence rate of the spatial median. A key distinction between the two estimators of $\S=\K$, namely $\hat{\K}$ and $\hat{\S}$, lies in the necessity to estimate the location parameter for $\hat{\S}$. This additional step introduces complexity to the proof of its convergence rate. Conversely, the theoretical analysis of $\hat{\K}$ is simplified by obviating the need to estimate the location parameter. However, this simplification comes at the cost of increased computational burden, as $\hat{\K}$ is a second-order U-statistic, whereas $\hat{\S}$ is only a first-order U-statistic. Ultimately, if $\frac{r^*(\S)}{n^\delta\log d} \to 0$, the second term will be of smaller order compared to the first term.

According to the Davis-Kahan inequality, we know that
$$
\left|\sin \angle\left(\boldsymbol{u}_1(\widehat{\mathbf{S}}), \boldsymbol{u}_1(\mathbf{S})\right)\right| \leq \frac{2}{\lambda_1(\mathbf{S})-\lambda_2(\mathbf{S})}\|\widehat{\mathbf{S}}-\mathbf{S}\|_2.
$$
So we can directly obtain the following corollary.

\begin{coro}\label{cor1}
Under the conditions of Theorem \ref{th1}, for any $\alpha>0$, there exist a positive constant $C_S$, such that, for sufficient large $n$ and $\delta\in (0,1)$,
we have, with probability larger than $1-\alpha$,
$$
\begin{aligned}
& \left|\sin \angle\left(\boldsymbol{u}_1(\widehat{\mathbf{S}}), \boldsymbol{u}_1(\mathbf{S})\right)\right| \\
& \quad \leq \frac{2 \lambda_1(\mathbf{S})}{\lambda_1(\mathbf{S})-\lambda_2(\mathbf{S})} \sqrt{\frac{4\left(r^*(\mathbf{S})+1\right)(\log d+\log (1 / \alpha))}{n}}+\frac{2C_S}{\lambda_1(\S)-\lambda_2(\S)}n^{-\frac{1}{2}(1+\delta)}
\end{aligned}
$$
\end{coro}

If $\lambda_1(\S)/\lambda_2(\S)$ is bounded by a constant, we need $r^*(\S)\log d/n \to 0$ and $r^*(\S)n^{-\frac{1}{2}(1+\delta)}\to 0$ to make $\u_1(\hat{\S})$ a consistent estimator of $\u_1(\S)$. If $\log (d)=o(n^{\frac{1}{2}(1-\delta)})$, we only need the assumption $r^*(\S)\log d/n \to 0$, which is consistent with the result in \cite{han2018eca}. According to Theorem 3.2 in \cite{han2018eca}, 
\begin{align*}
r^*(\S)\le \left(r^*(\boldsymbol{\Sigma})+4 r^{* *}(\boldsymbol{\Sigma}) \sqrt{\log d}+8 \log d\right)\left(1-\sqrt{3} d^{-2}\right)^{-1}
\end{align*}
where $r^{* *}(\boldsymbol{\Sigma}):=\|\boldsymbol{\Sigma}\|_F / \lambda_1(\boldsymbol{\Sigma}) \leq \sqrt{d}$ is the ``second-order" effective rank of the matrix $\boldsymbol{\Sigma}$. Additionally, if { $\|\bms\|_F\log d=o(1)\tr(\bms)$}, we have $\lambda_j(\S)\asymp\lambda_j(\bms)/\tr(\bms)$ { when $d\rightarrow \infty$.} So,
$$
\frac{\lambda_1(\mathbf{S})}{\lambda_1(\mathbf{S})-\lambda_2(\mathbf{S})} \asymp \frac{\lambda_1(\boldsymbol{\Sigma})}{\lambda_1(\boldsymbol{\Sigma})-\lambda_2(\boldsymbol{\Sigma})},
$$
{ as $d\rightarrow \infty.$}
So, we can directly bound $\|\hat \S-\S\|_2$ and $\left|\sin \angle\left(\boldsymbol{u}_1(\widehat{\mathbf{S}}), \boldsymbol{u}_1(\mathbf{S})\right)\right|$ using $\bms$.

We observe that Theorem \ref{th1} can also facilitate the quantification of the subspace estimation error through a variant of the Davis-Kahan inequality. Specifically, let $\mathcal{P}^m(\widehat{\mathbf{S}})$ and $\mathcal{P}^m(\mathbf{S})$ denote the projection matrices that map onto the subspaces spanned by the $m$ leading eigenvectors of $\widehat{\mathbf{S}}$ and $\mathbf{S}$, respectively. By invoking Lemma 4.2 from \cite{VuLei2013}, we obtain the inequality:
$$
\left\|\mathcal{P}^m(\widehat{\mathbf{S}})-\mathcal{P}^m(\mathbf{S})\right\|_{\mathrm{F}} \leq \frac{2 \sqrt{2 m}}{\lambda_m(\mathbf{S})-\lambda_{m+1}(\mathbf{S})}\|\widehat{\mathbf{S}}-\mathbf{S}\|_2,
$$
which allows us to control $\left\|\mathcal{P}^m(\widehat{\mathbf{S}})-\mathcal{P}^m(\mathbf{S})\right\|_{\mathrm{F}}$ using a similar rationale as employed in Corollary \ref{cor1}.

\section{Sparse SPCA: Sparse Setting}
\subsection{Combinatoric Program}
In this section, we consider sparse settings: $\lambda_1(\bms)$ is distinct and $\|\u_1(\bms)\|_0\le s<d\wedge n$. For any matrix $\mathbf{M} \in \mathbb{R}^{d \times d}$, we define the best $s$-sparse vector approximating $\boldsymbol{u}_1(\mathbf{M})$ as
\begin{align}\label{sp}
\boldsymbol{u}_{1, s}(\mathbf{M}):=\underset{\|\boldsymbol{v}\|_0 \leq s,\|\boldsymbol{v}\|_2 \leq 1}{\arg \max }\left|\boldsymbol{v}^T \mathbf{M} \boldsymbol{v}\right|
\end{align}

We propose to estimate $\boldsymbol{u}_1(\boldsymbol{\Sigma})=\boldsymbol{u}_1(\mathbf{S})$ via a combinatoric program:

{\it Sparse SPCA estimator (SSPCA) via a combinatoric program : $\boldsymbol{u}_{1, s}(\widehat{\mathbf{S}})$.}

Similarly, to evaluate the performance of SSPCA, we first study the approximation error $\|\hat\S-\S\|_{2,s}$.

\begin{theorem}\label{th2}
Let $\boldsymbol{X}_1, \ldots, \boldsymbol{X}_n$ be $n$ observations of $\boldsymbol{X} \sim$ $E C_d(\boldsymbol{\mu}, \boldsymbol{\Sigma}, \xi)$,   when $(s \log (e d / s)+\log (1 / \alpha)) / n \rightarrow 0$, with probability at least $1-3 \alpha$, we have
$$
\begin{aligned}
\|\widehat{\mathbf{S}}-\mathbf{S}\|_{2, s} \leq &  C_0 \left(\sup _{\boldsymbol{v} \in \mathbb{S}^{d-1}} 2\left\|\boldsymbol{v}^T U(\boldsymbol{X})\right\|_{\psi_2}^2+\|\mathbf{S}\|_2\right)\sqrt{\frac{s(3+\log (d / s))+\log (1 / \alpha)}{n}}+C_1 \left(\frac{nd}{s}\right)^{-\frac{1}{2}(1+\delta)}
\end{aligned}
$$
for some absolute constants $C_0,C_1 > 0$ and $\delta \in (0,1)$.
Specially, if $\operatorname{rank}(\boldsymbol{\Sigma})=q$ and $\left\|\boldsymbol{u}_1(\boldsymbol{\Sigma})\right\|_0 \leq s$, we have
$$
\begin{aligned}
\|\widehat{\mathbf{S}}-\mathbf{S}\|_{2, s} \leq & C_0\left\{\left(\frac{4 \lambda_1(\mathbf{\Sigma})}{q \lambda_q(\boldsymbol{\Sigma})} \wedge 1\right)+\lambda_1(\mathbf{S})\right\}  \sqrt{\frac{s(3+\log (d / s))+\log (1 / \alpha)}{n}}+C_1 \left(\frac{nd}{s}\right)^{-\frac{1}{2}(1+\delta)}
\end{aligned}
$$
\end{theorem}
The first term of the approximation error of $\|\widehat{\mathbf{S}}-\mathbf{S}\|_{2, s}$ is the same as $\|\widehat{\mathbf{K}}-\mathbf{K}\|_{2, s}$ by Theorem 4.1 and 4.2 in \cite{han2018eca}. Similar to Theorem \ref{th1}, the second term arises from the convergence rate of the spatial median. Under some special cases, such as condition number controlled \citep{Bickel2008}, spike covariance model \citep{Johnstone2009}, multi-factor model \citep{Fan2008}, \cite{han2018eca} showed that $\sup_{\v}\left\|\boldsymbol{v}^T U(\boldsymbol{X})\right\|_{\psi_2}^2$ is of the same order as $\lambda_1(\mathbf{S})$. So the first term is $O_P\left(\lambda_1(\mathbf{S}) \sqrt{s \log (e d / s) / n}\right)$. Then, if $\frac{r^*(\S)s^{\delta/2}}{n^{\delta/2}d^{\frac{1+\delta}{2}}\sqrt{\log (ed/s)}}\to 0$, the second term is a smaller order than the first term.

By Davis-Kahan type inequality provided in \cite{VuLei2012}, we have
$$
\left|\sin \angle\left(\boldsymbol{u}_{1, s}(\widehat{\mathbf{S}}), \boldsymbol{u}_{1, s}(\mathbf{S})\right)\right| \leq \frac{2}{\lambda_1(\mathbf{S})-\lambda_2(\mathbf{S})}\|\widehat{\mathbf{S}}-\mathbf{S}\|_{2,2 s}.
$$
So we can directly obtain the following result.

\begin{coro}\label{cor2}
Under the condition of Theorem \ref{th2}, if we have $(s \log (e d / s)+\log (1 / \alpha)) / n \rightarrow 0$, for $n$ sufficiently large, with probability larger than $1-2 \alpha$,
$$
\begin{aligned}
\left|\sin \angle\left(\boldsymbol{u}_{1, s}(\widehat{\mathbf{S}}), \boldsymbol{u}_{1, s}(\mathbf{S})\right)\right| \leq & \frac{2 C_0\left(4 \lambda_1(\boldsymbol{\Sigma}) / q \lambda_q(\boldsymbol{\Sigma}) \wedge 1+\lambda_1(\mathbf{S})\right)}{\lambda_1(\mathbf{S})-\lambda_2(\mathbf{S})} \sqrt{\frac{2 s(3+\log (d / 2 s))+\log (1 / \alpha)}{n}}\\
&+\frac{2C_1}{\lambda_1(\mathbf{S})-\lambda_2(\mathbf{S})}\left(\frac{nd}{s}\right)^{-\frac{1}{2}(1+\delta)}
\end{aligned}
$$
\end{coro}
By \cite{WangHanLiu2013}, we have
\begin{align*}
& \left\|\mathbf{U}_{m, s}(\widehat{\mathbf{S}}) \mathbf{U}_{m, s}(\widehat{\mathbf{S}})^T-\mathbf{U}_{m, s}(\mathbf{S}) \mathbf{U}_{m, s}(\mathbf{S})^T\right\|_{\mathrm{F}} \leq \frac{2 \sqrt{2 m}}{\lambda_m(\mathbf{S})-\lambda_{m+1}(\mathbf{S})} \cdot\|\widehat{\mathbf{S}}-\mathbf{S}\|_{2,2 m s}
\end{align*}
where 
$$
\mathbf{U}_{m, s}(\mathbf{M}):=\underset{\mathbf{V} \in \mathbb{R}^{d \times m}}{\arg \max }\left\langle\mathbf{M}, \mathbf{V} \mathbf{V}^T\right\rangle \text {, subject to } \sum_{j=1}^d \mathbb{I}\left(\mathbf{V}_{j *} \neq 0\right) \leq s,
$$
where $V_{j *}$ is the $j$ th row of $\mathbf{M}$ and the indicator function returns 0 if and only if $\mathbf{V}_{j *}=\mathbf{0}$. Then, the results obtained in Theorem \ref{th2} can also used to bound the approximation error of the principal subspace estimation.

\subsection{Computationally Efficient program}
We adopt the truncated power algorithm proposed by \cite{YuanZhang2013} to solve the optimization problem (\ref{sp}). For any vector $\boldsymbol{v} \in \mathbb{R}^d$ and an index set $J \subset\{1, \ldots, d\}$, we define the truncation function $\operatorname{TRC}(\cdot, \cdot)$ to be
$
\operatorname{TRC}(\boldsymbol{v}, J):=\left(v_1 \cdot \mathbb{I}(1 \in J), \ldots, v_d \cdot \mathbb{I}(d \in J)\right)^T
$
where $\mathbb{I}(\cdot)$ is the indicator function. Algorithm 1 shows the detail procedures of our proposed SSPCA procedure.

\begin{algorithm}[ht]
\caption{Sparse Spatial-sign bases Principal Component Analysis (SSPCA)}
\begin{algorithmic}[1]
\Require{Matrix $\hat{\mathbf{S}}$, sparsity level $k$, convergence threshold $\epsilon$}
\Ensure{$\hat{u}_{1, k}(\hat{\mathbf{S}})$}
\State The initial parameter $\boldsymbol{v}^{(0)}$.
\Repeat
    \State $t \leftarrow t+1$.
    \State Compute $\boldsymbol{W}_t \leftarrow \hat{\mathbf{S}} \boldsymbol{v}^{(t-1)}$.
    \If{$\left\|\boldsymbol{W}_t\right\|_0 \leq k$}
        \State $\boldsymbol{v}^{(t)} \leftarrow \boldsymbol{W}_t /\left\|\boldsymbol{W}_t\right\|_2$.
    \Else
        \State Let $A_t$ be the indices of the elements in $\boldsymbol{W}_t$ with the $k$ largest absolute values.
        \State $\boldsymbol{v}^{(t)} \leftarrow \operatorname{TRC}\left(\boldsymbol{W}_t, A_t\right) /\left\|\operatorname{TRC}\left(\boldsymbol{W}_t, A_t\right)\right\|_2$.
    \EndIf
\Until{$\left\|v^{(t)}-v^{(t-1)}\right\|_2 \leq \epsilon$}
\State $\hat{\u}_{1, k}(\hat{\mathbf{S}}) \leftarrow v^{(t)}$.
\end{algorithmic}
\end{algorithm}

The following theorem show the consistency of Algorithm 1, which is a directly result of Theorem 4 in \cite{YuanZhang2013}. So we omit the detailed proof here.
\begin{theorem}
Suppose $\left\|\boldsymbol{v}^{(0)}\right\|_0 \leq s$ and $\left|\left(\boldsymbol{v}^{(0)}\right)^T \boldsymbol{u}_1(\mathbf{S})\right|$ is lower bounded by a positive constant $C_3$. Accordingly under the condition of Theorem 4 by \cite{YuanZhang2013}, for $k \geq s$, we have
$$
\left|\sin \angle\left(\widehat{\boldsymbol{u}}_{1, k}(\widehat{\mathbf{S}}), \boldsymbol{u}_1(\mathbf{S})\right)\right|=O_P\left(\sqrt{\frac{(k+s) \log d}{n}}\right).
$$
\end{theorem}

In practical applications, we have observed that the leading eigenvector of $\hat{\S}$ exhibits excellent performance as an initial parameter. Therefore, we adopt this simpler initial estimator in our paper. Similar to the approach by \cite{han2018eca}, the initial parameter $\bm v^{(0)}$ can be estimated using the Fantope Projection method proposed by \cite{VuChoLeiRohe2013}. We introduce this method in the appendix.

\subsection{Tuning parameter selection}

The tuning parameter $k$ in Algorithm 1 plays a crucial role in the performance of sparse PCA. A large value of $k$ may result in the inclusion of numerous unimportant parameters, while a small value of $k$ may lead to significant bias. One potential approach to selecting $k$ is to utilize the criterion proposed by \cite{YuanZhang2013}, which involves choosing the value of $k$ that maximizes $\left(\widehat{\boldsymbol{u}}_{1, k}(\widehat{\mathbf{S}})\right)^T \cdot \widehat{\mathbf{S}}_{\text {val }} \cdot \widehat{\boldsymbol{u}}_{1, k}(\widehat{\mathbf{S}})$, where $\widehat{\mathbf{S}}_{\text {val }}$ represents an independent empirical spatial-sign covariance matrix calculated from a separate sample set of the data. \cite{YuanZhang2013} demonstrated that this heuristic approach performs well in practical applications. However, in situations where an independent sample set is not available, we recommend using the sample-split method as an alternative approach.

For each $k$, we randomly split the sample into two sets, denote the corresponding sample spatial-sign covariance matrix of each sample as $\hat \S_{l}^{(1)}$, $\hat \S_{l}^{(2)}$, respectively. Then, we calculate
\begin{align}\label{sk}
\hat{k}=\arg\max_{1\le k\le K} \frac{1}{B}\sum_{l=1}^B \left(\widehat{\boldsymbol{u}}_{1, k}(\widehat{\mathbf{S}_l^{(1)}})\right)^T \cdot \widehat{\mathbf{S}}_{l}^{(2)} \cdot \widehat{\boldsymbol{u}}_{1, k}(\widehat{\mathbf{S}_l^{(1)}})
\end{align}


In the above discussion, we only consider estimate the leading eigenvector. To estimate more than one leading eigenvectors, we exploit the deflation method proposed by \cite{mackey2008}. That is, we obtain multiple component estimates by taking the $r$-th component estimate $\hat{\bm v}_r$ from input matrix $\S_r$, and then re-running the method with the deflated input matrix: $\S_{r+1}=\left(\I-\hat{\bm v}_r \hat{\bm v}_r^T\right) \S_r\left(\I-\hat{\bm v}_r \hat{\bm v}_r^T\right)$. The resulting $m$-dimensional principal subspace estimate is the span of $\hat{\bm v}_1, \ldots, \hat{\bm v}_m$.

\section{Simulation}
\subsection{Estimating Leading Eigenvector}
\begin{figure}[htbp]
\centering
\caption{Curves of averaged distances between the estimates and true parameters with different number of selected features. \label{figad1}}
\subfloat[$n=100,d=100$]{\includegraphics[width=0.85\textwidth]{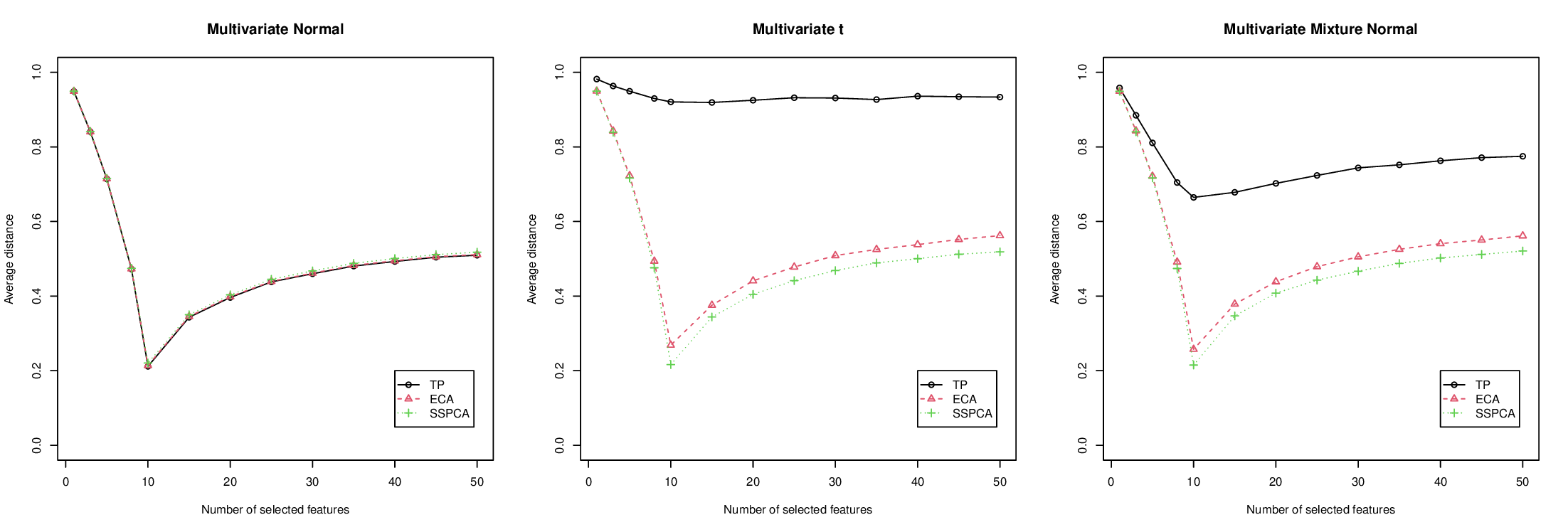}}\\
\subfloat[$n=100,d=200$]{\includegraphics[width=0.85\textwidth]{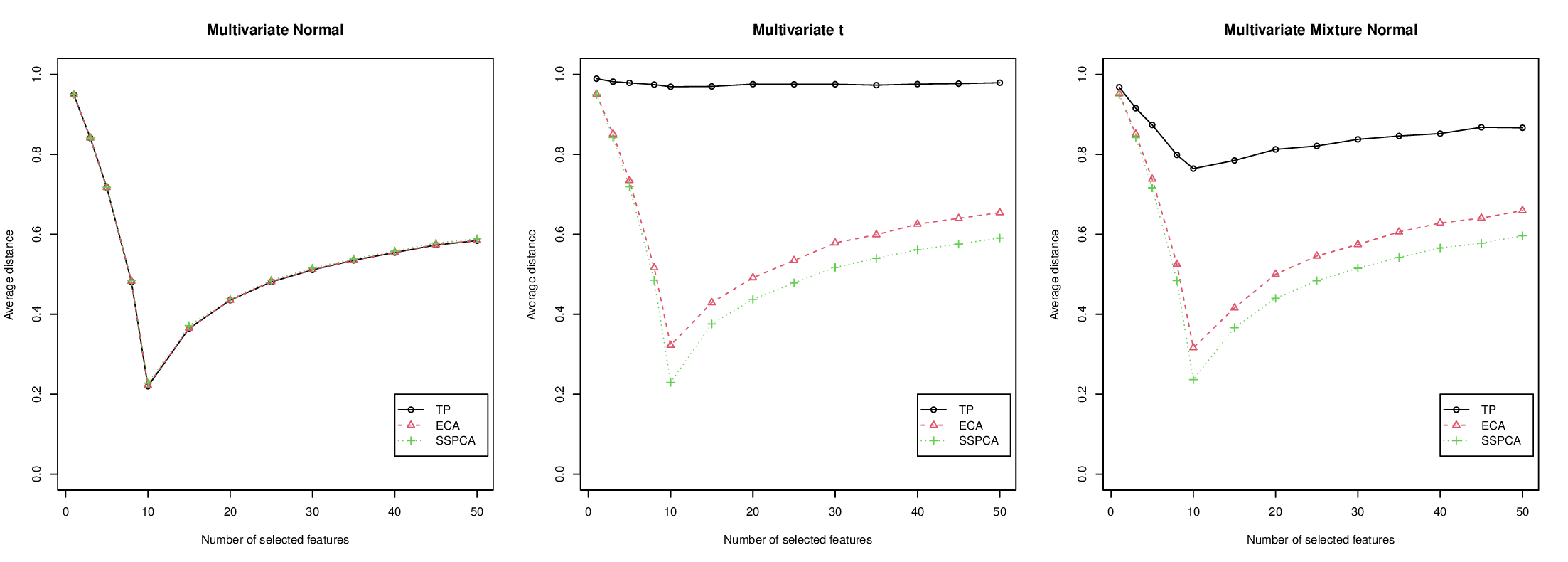}}\\
\subfloat[$n=100,d=300$]{\includegraphics[width=0.85\textwidth]{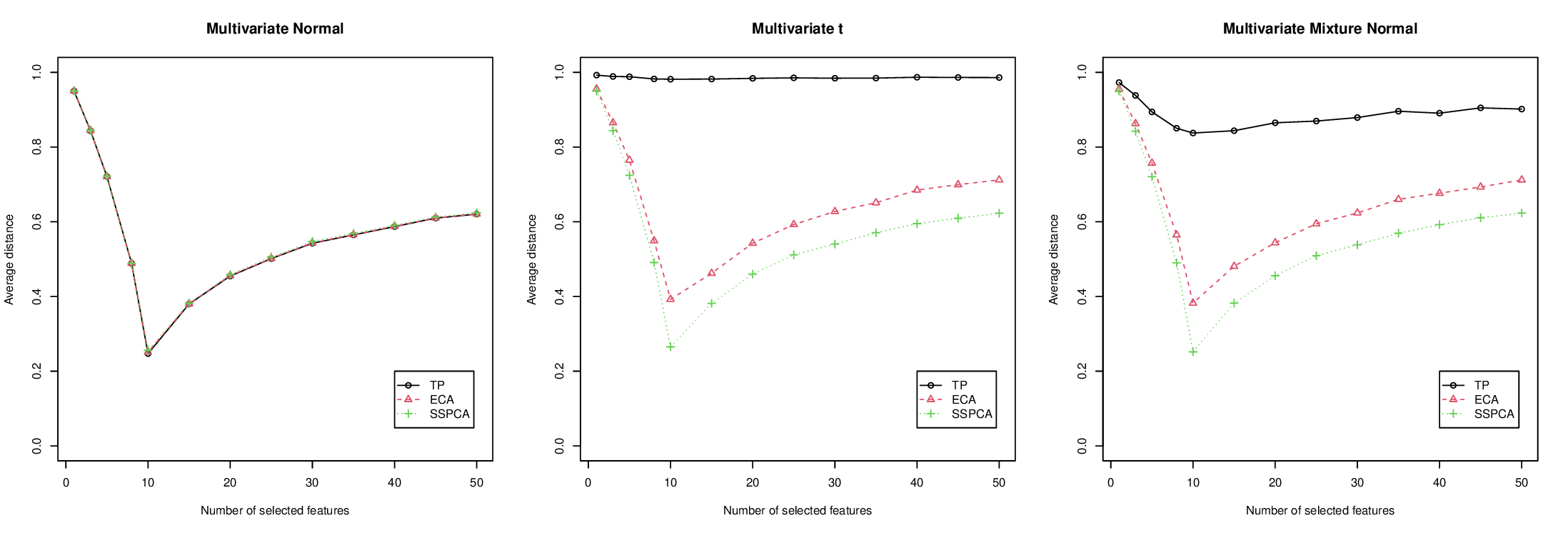}}
\end{figure}

We first consider estimating the leading eigenvector of the covariance matrix ${\bf \Sigma}$. We consider the similar model for $\bf \Sigma$ as \cite{han2018eca}, i.e.
\begin{align*}
\boldsymbol{\Sigma}=\sum_{j=1}^m\left(\omega_j-\omega_d\right) \boldsymbol{v}_j \boldsymbol{v}_j^T+\omega_d \mathbf{I}_d
\end{align*}
where $\omega_1>\omega_2>\omega_3=$ $\cdots=\omega_d$ be the eigenvalues and $\boldsymbol{v}_1, \ldots, \boldsymbol{v}_d$ be the eigenvectors of $\boldsymbol{\Sigma}$ with $\boldsymbol{v}_j:=\left(v_{j 1}, \ldots, v_{j d}\right)^T$. The top $m$ leading eigenvectors $\boldsymbol{v}_1, \ldots, \boldsymbol{v}_m$ of $\boldsymbol{\Sigma}$ are specified to be sparse such that $s_j:=\left\|\boldsymbol{v}_j\right\|_0$ is small and
$$
v_{j k}= \begin{cases}1 / \sqrt{s_j}, & 1+\sum_{i=1}^{j-1} s_i \leq k \leq \sum_{i=1}^j s_i \\ 0, & \text { otherwise. }\end{cases}
$$
In this subsection, we set $m=2$ and $\omega_1=5,\omega_2=3,\omega_3=\cdots=\omega_p=1$.
We consider the following three different elliptical distributions:
\begin{itemize}
\item[(I)] Multivariate normal distribution. $\bm X\sim N({ \bm 0},{\bf \Sigma})$.
\item[(II)] Multivariate $t$-distribution $t_{N,3}$.   $\bm X$'s are generated from standardized $t_{N,3}/\sqrt{3}$ with mean zero and scatter matrix ${\bf \Sigma}$.
\item[(III)] Multivariate mixture normal distribution $\mbox{MN}_{N,\kappa,9}$. $\bm X$'s are generated from standardized  $[\kappa
N(\bm 0,{\bf \Sigma})+(1-\kappa)N(\bm 0,9{\bf \Sigma})]/\sqrt{\kappa+9(1-\kappa)}$, denoted
by $\mbox{MN}_{N,\gamma,9}$. $\kappa$ is chosen to be 0.8.
\end{itemize}
All the simulation results are based on 1000 replication. Figure \ref{figad1} plots the averaged distances between the estimate $\widehat{\boldsymbol{v}}_1$ and $\boldsymbol{v}_1$, defined as $\left|\sin \angle\left(\widehat{\boldsymbol{v}}_1, \boldsymbol{v}_1\right)\right|$, against the number of estimated nonzero entries (defined as $\left.\left\|\widehat{\boldsymbol{v}}_1\right\|_0\right)$, for three different methods:
\begin{itemize}
\item TP: Sparse PCA method on the Pearson's sample covariance matrix (Yuan and Zhang, 2013);
\item ECA: Elliptical component analysis based on the multivariate kendall's tau matrix \citep{han2018eca}.
\item SSPCA: Sparse spatial-sign based Principal component analysis.
\end{itemize}
In this case, we set $n=100$ and varied $d$ to be $100, 200, 300$. Our findings indicate that SSPCA consistently outperforms ECA and TP in estimation accuracy. This result underscores the effectiveness of SSPCA in handling high-dimensional data with potential sparsity. SSPCA's ability to accurately estimate the principal components in high-dimensional settings is a crucial advantage, as many modern datasets are characterized by a large number of features. Furthermore, when the data are indeed normally distributed, we observed no significant difference in performance between SSPCA, ECA, and TP. This observation suggests that SSPCA is a reliable alternative to sparse PCA within the elliptical family of distributions. The fact that SSPCA performs comparably to other methods in the case of normally distributed data, while also excelling in high-dimensional and sparse settings, demonstrates its versatility and robustness. Overall, these findings highlight the potential of SSPCA as a powerful tool for analyzing high-dimensional data in a variety of contexts.

\begin{figure}[htbp]
\centering
\caption{Curves of averaged distances between the estimates and true parameters with varying number of dimensions and sample size. \label{figad2}}
\subfloat[$s=5$]{\includegraphics[width=0.85\textwidth]{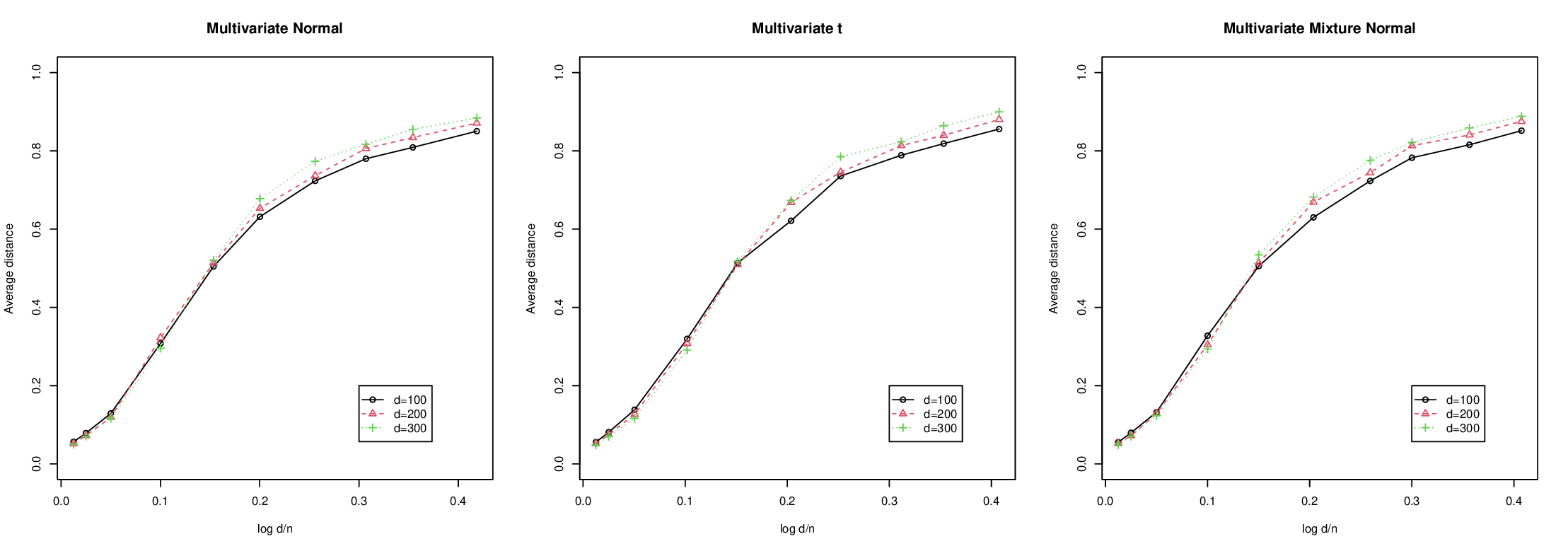}}\\
\subfloat[$s=10$]{\includegraphics[width=0.85\textwidth]{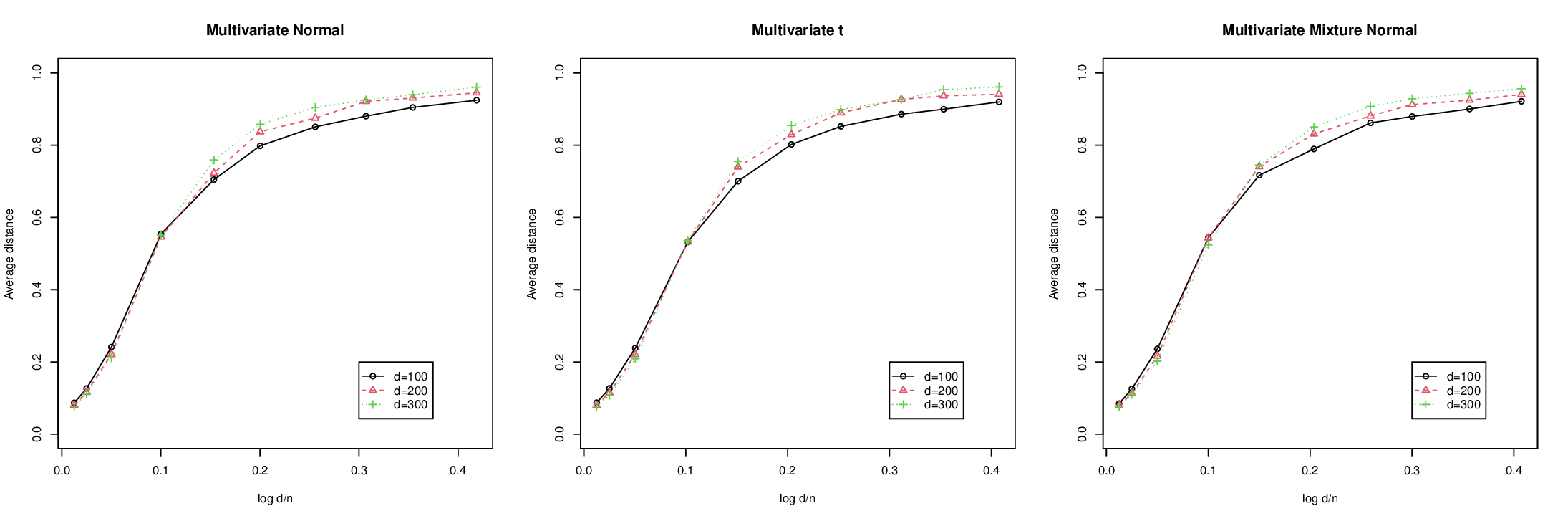}}\\
\subfloat[$s=20$]{\includegraphics[width=0.85\textwidth]{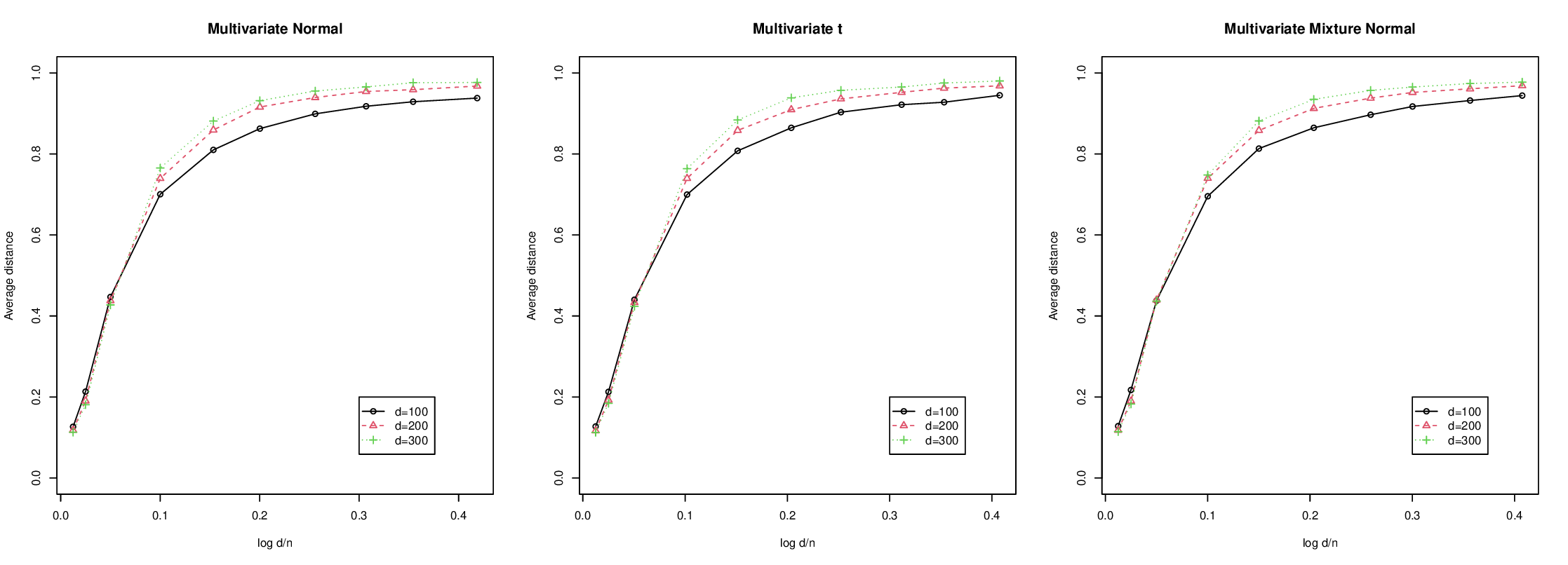}}
\end{figure}

To evaluate the dependence of the estimation accuracy of the SSPCA estimator on the triplet $(n,d,s)$, we conducted experiments with varying values of $d$, $s$, and sample size $n$. Specifically, we considered $d=100, 200, 300$, $s_1=5, 10, 20$, and varying sample sizes $n$. The results, presented in Figure \ref{figad2}, show the curves of averaged distances between the estimates and true parameters. In these experiments, we set the number of selected features equal to the true parameter $s$. Our findings indicate that the averaged distance between $\bm v_1$ and $\hat{\bm v}_1$ approaches zero as the sample size increases, which demonstrates the consistency of our proposed SSPCA methods. This consistency is an important characteristic of any estimator, as it indicates that the estimates produced by the method will be increasingly accurate as more data is available. Additionally, we observed that all the curves in Figure \ref{figad2} almost overlap with each other when the average distances are plotted against $\log d/n$. This observation is consistent with the results presented in Corollary \ref{cor1}, which suggests that the effective sample size is $n/\log d$ when controlling the prediction accuracy of the eigenvectors. This finding highlights the importance of considering the relationship between $n$ and $d$ when evaluating the performance of SSPCA. Specifically, it suggests that as the dimension $d$ increases, the sample size $n$ must also increase in order to maintain a given level of prediction accuracy.

\begin{figure}[htbp]
\centering
\caption{Average computation time of SSPCA and ECA with $d=100$ and varying sample sizes. \label{figad3}}
{\includegraphics[width=0.85\textwidth]{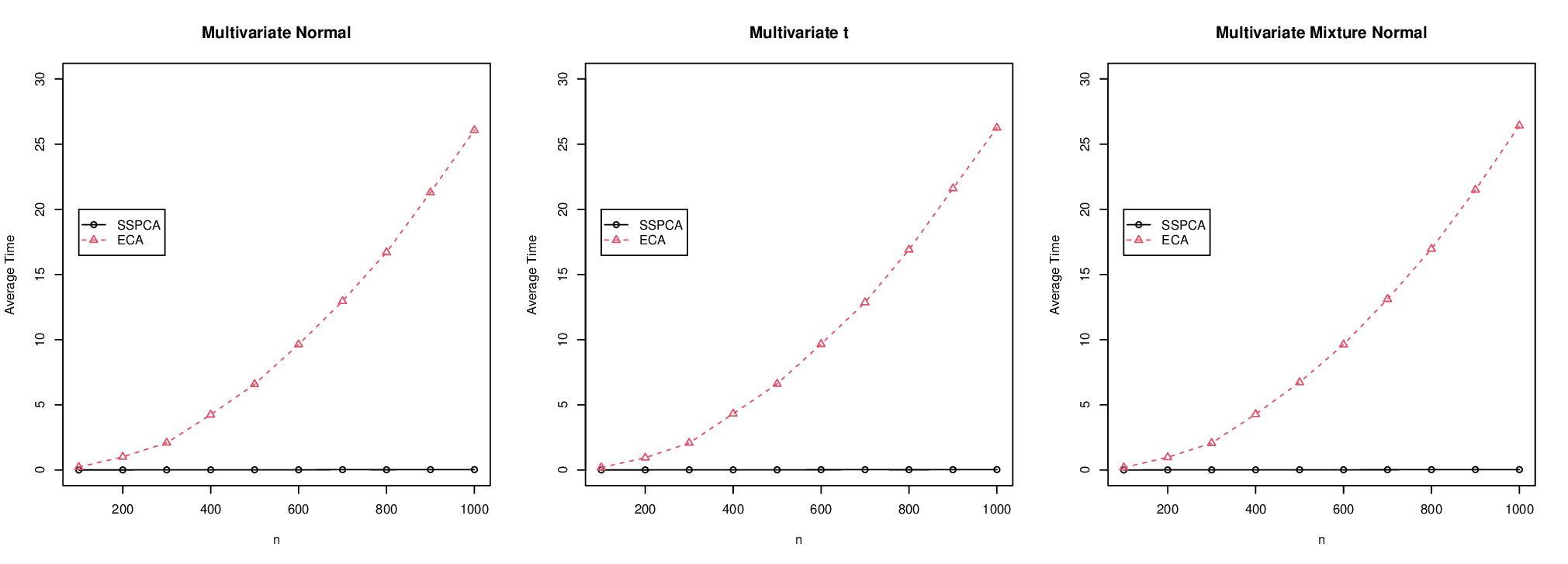}}
\end{figure}

To demonstrate the computational efficiency of our proposed SSPCA method, we conducted experiments with $d=100$ and varying sample sizes. The results, presented in Figure \ref{figad3}, show the average computation time for both SSPCA and the existing method, ECA. Our findings indicate that the average computation time of SSPCA grows linearly with the sample size, whereas the computation time of ECA grows quadratically with the sample size. This observation highlights a significant advantage of SSPCA, particularly when dealing with large sample sizes. The linear growth in computation time suggests that SSPCA is able to efficiently handle increasing amounts of data, making it a preferable choice for large-scale datasets. In contrast, the quadratic growth of ECA's computation time indicates that it may become impractical for large sample sizes due to the significantly increased computational burden. Therefore, our proposed SSPCA method offers a computationally efficient solution for analyzing large datasets, making it a valuable tool for researchers and practitioners working with big data.

\begin{figure}[htbp]
\centering
\caption{Histogram of estimator of the number of selected features with $n=400,d=300$. \label{figss}}
\subfloat[Multivariate Normal]{\includegraphics[width=0.85\textwidth]{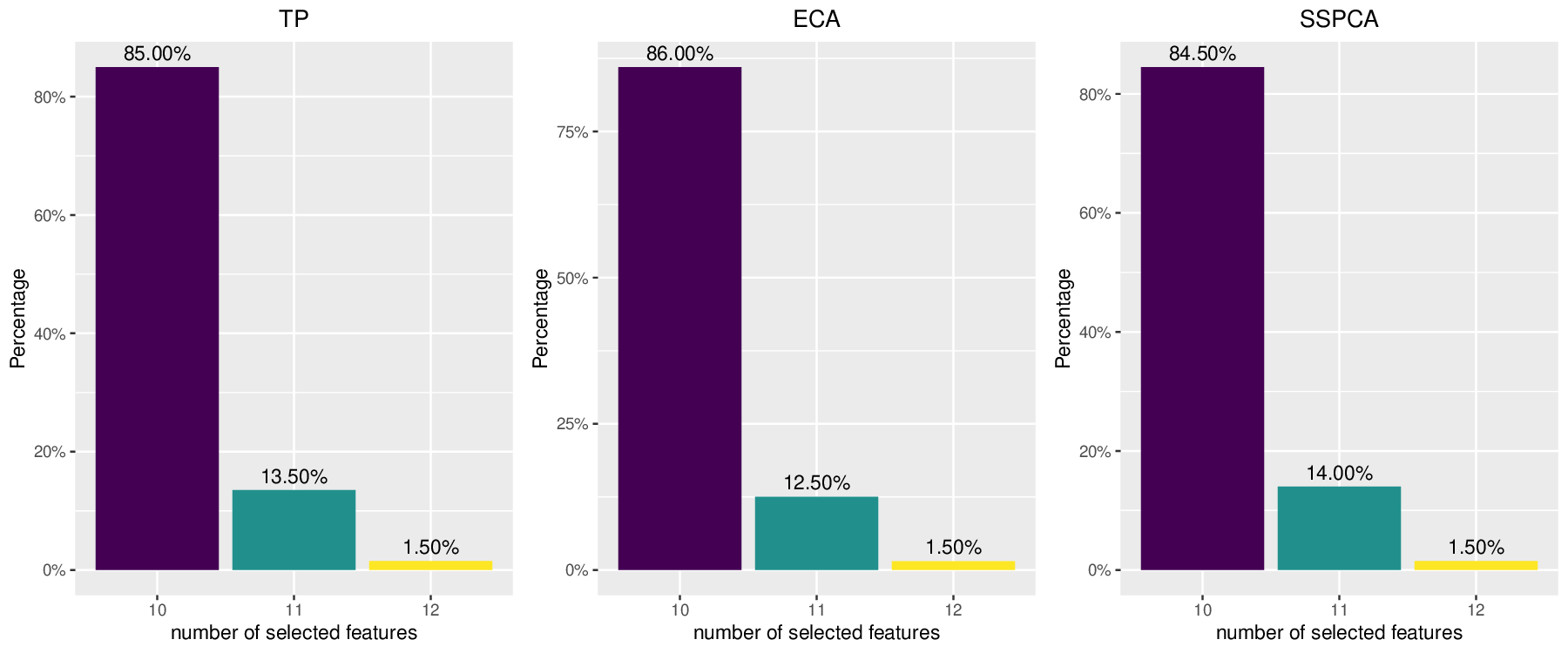}}\\
\subfloat[Multivariate t]{\includegraphics[width=0.85\textwidth]{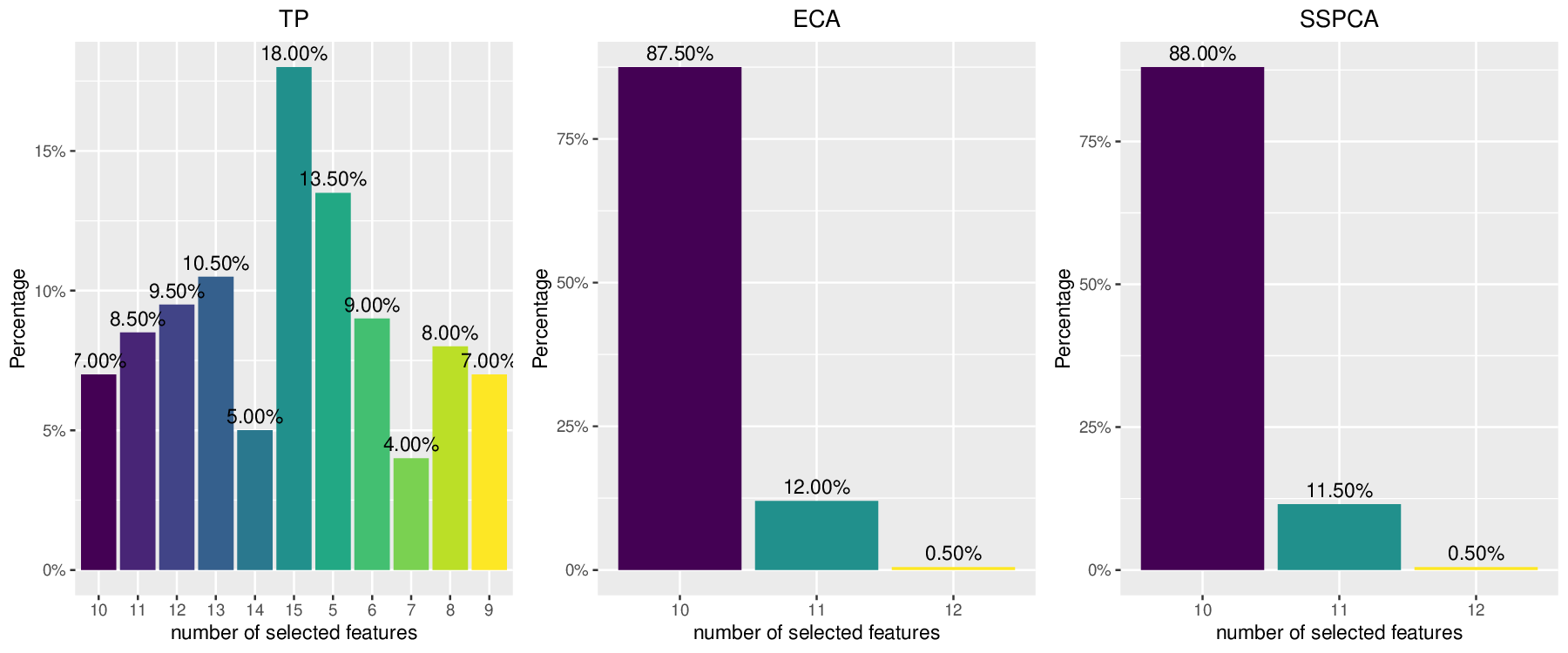}}\\
\subfloat[Multivariate Mixture Normal]{\includegraphics[width=0.85\textwidth]{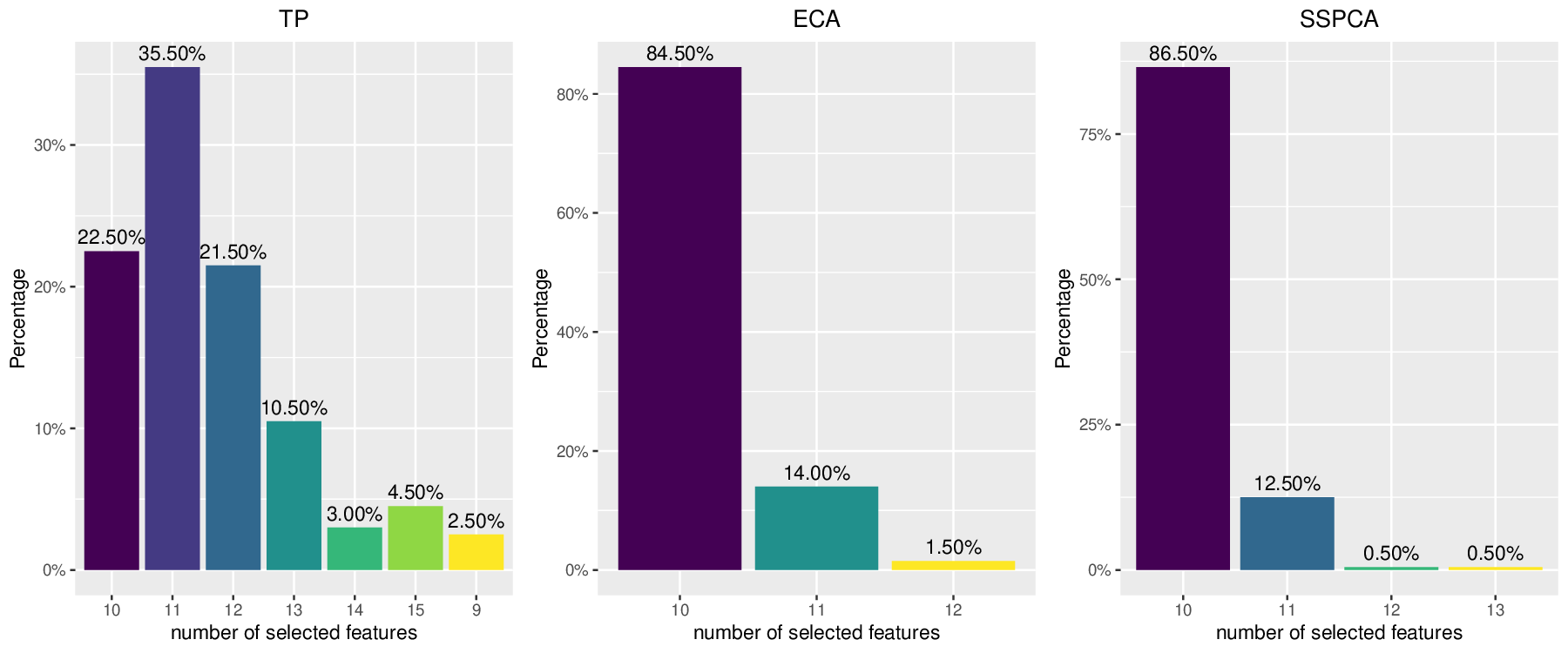}}
\end{figure}

Next, we evaluate the estimation accuracy of the number of selected features. Table \ref{t2} reports the average distances between the estimates and true parameters of the leading eigenvector, comparing the estimated number of selected features $\hat{s}$ with the true number of selected features $s$. We observe that the average distance with the estimated $\hat{s}$ is slightly larger than the oracle estimator with the true $s$. As the sample size increases, the estimation of $s$ improves, leading to a smaller average distance $\left|\sin \angle\left(\widehat{\boldsymbol{v}}_1, \boldsymbol{v}_1\right)\right|$ between $\hat{s}$ and $s$. Figure \ref{figss} shows the histogram of the estimated number of selected features with $n=400$ and $d=300$. Our findings indicate that both ECA and SSPCA consistently estimate the number of selected features, whereas TP does not perform well with heavy-tailed distributions.

\begin{table}[!ht]
\begin{center}
\caption{\label{t2} The averaged distances between the estimates and true parameters of the leading eigenvector with estimated number of selected features $\hat{s}$ and the true number of selected features $s$.}
                     \vspace{0.5cm}
                     \renewcommand{\arraystretch}{0.8}
                     \setlength{\tabcolsep}{3pt}{
\begin{tabular}{c|cc|cc|cc|cc|cc|cc}
\hline \hline
 $T$ & \multicolumn{6}{c}{{$n=200$}} & \multicolumn{6}{c}{{$n=400$}}\\ \hline
Distributions &\multicolumn{2}{c}{(I)}&\multicolumn{2}{c}{(II)}&\multicolumn{2}{c}{(III)} &\multicolumn{2}{c}{(I)}&\multicolumn{2}{c}{(II)}&\multicolumn{2}{c}{(III)}\\ \hline
 & $\hat{s}$&$s$& $\hat{s}$&$s$& $\hat{s}$&$s$& $\hat{s}$&$s$& $\hat{s}$&$s$& $\hat{s}$&$s$\\ \hline
 &\multicolumn{12}{c}{$d=100$}\\ \hline
TP   &0.138&0.118&0.891&0.886&0.365&0.314&0.091&0.082&0.821&0.812&0.181&0.139\\
ECA  &0.139&0.118&0.169&0.133&0.163&0.135&0.091&0.083&0.096&0.088&0.095&0.087\\
SSPCA&0.141&0.121&0.140&0.119&0.148&0.125&0.091&0.083&0.090&0.082&0.094&0.085\\ \hline
 &\multicolumn{12}{c}{$d=200$}\\ \hline
TP   &0.141&0.113&0.938&0.936&0.441&0.401&0.091&0.083&0.923&0.917&0.208&0.146\\
ECA  &0.137&0.115&0.170&0.127&0.169&0.126&0.089&0.082&0.093&0.085&0.090&0.086\\
SSPCA&0.141&0.115&0.139&0.117&0.141&0.116&0.087&0.081&0.088&0.080&0.086&0.081\\ \hline
 &\multicolumn{12}{c}{$d=300$}\\ \hline
TP   &0.142&0.117&0.973&0.977&0.483&0.452&0.087&0.081&0.967&0.967&0.218&0.163\\
ECA  &0.142&0.117&0.185&0.132&0.192&0.134&0.086&0.081&0.095&0.090&0.094&0.087\\
SSPCA&0.142&0.118&0.145&0.118&0.150&0.120&0.087&0.081&0.089&0.084&0.088&0.083\\
\hline
\hline
\end{tabular}}
\end{center}
\end{table}

\subsection{Estimating Top $m$ Leading Eigenvector}
\begin{figure}[htbp]
\centering
\caption{Curves of averaged distances between the estimates and true parameters with different number of selected features on estimating top $m$ leading eigenvectors. \label{figadm}}
\subfloat[$n=50,p=100$]{\includegraphics[width=0.85\textwidth]{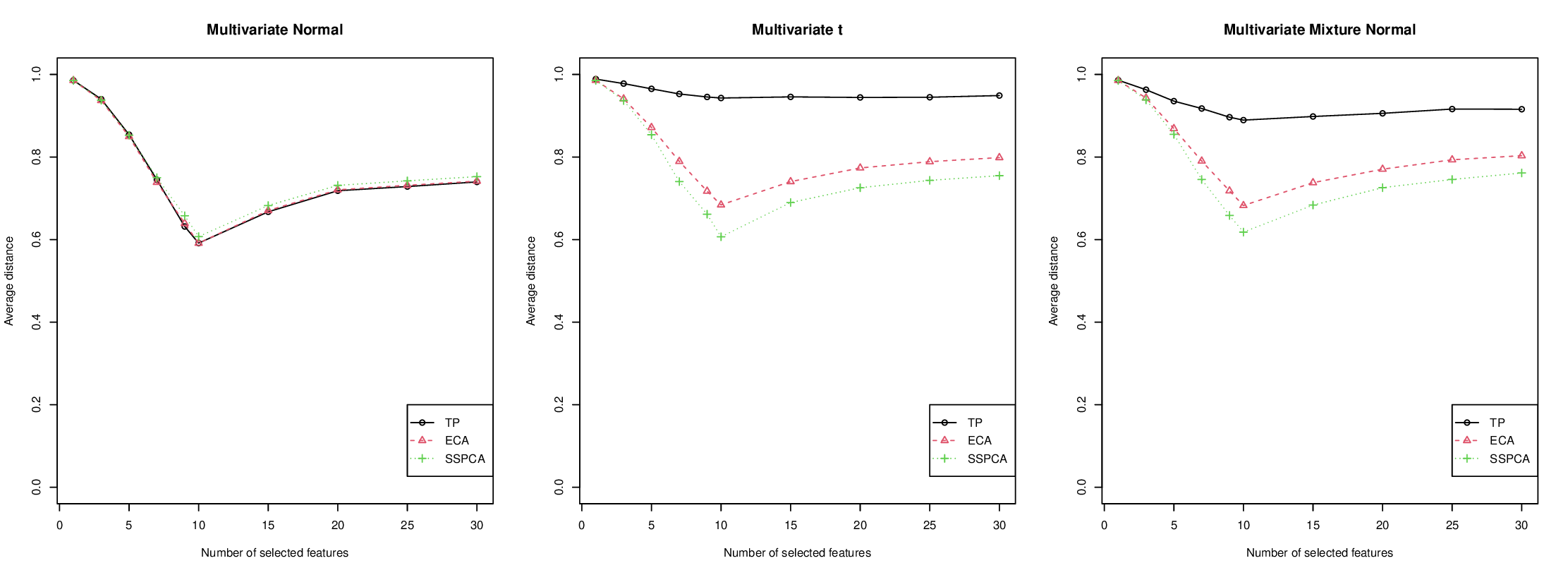}}\\
\subfloat[$n=100,p=100$]{\includegraphics[width=0.85\textwidth]{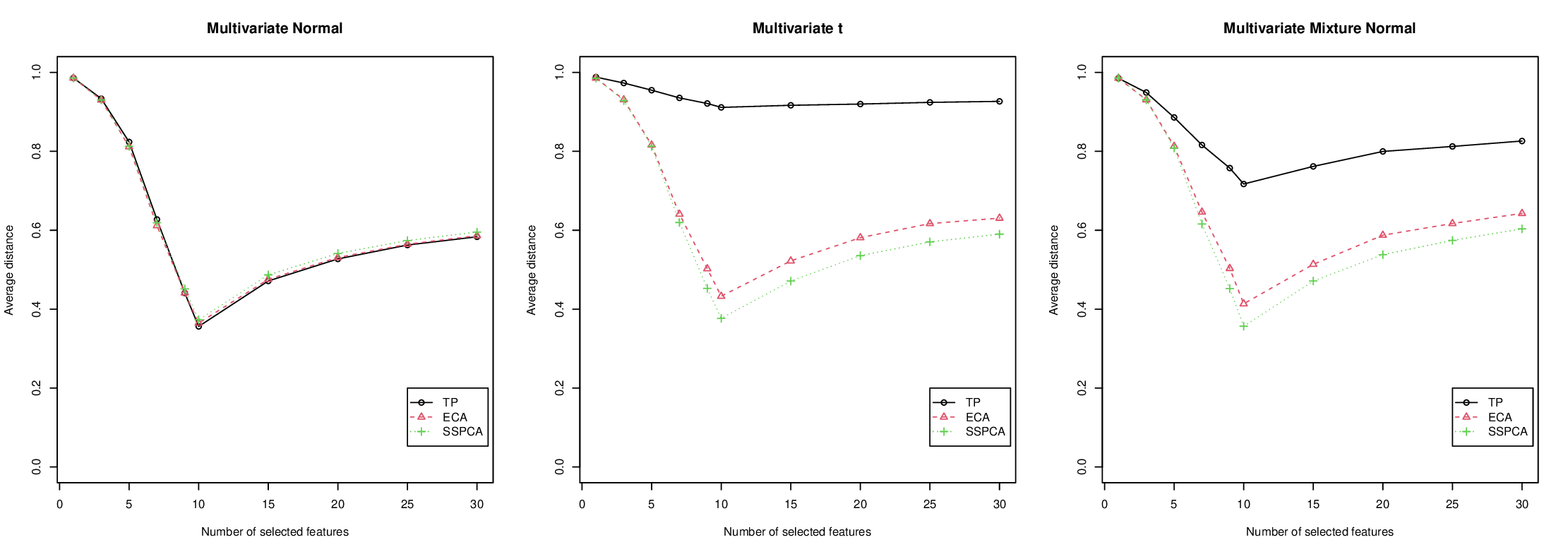}}\\
\subfloat[$n=200,p=100$]{\includegraphics[width=0.85\textwidth]{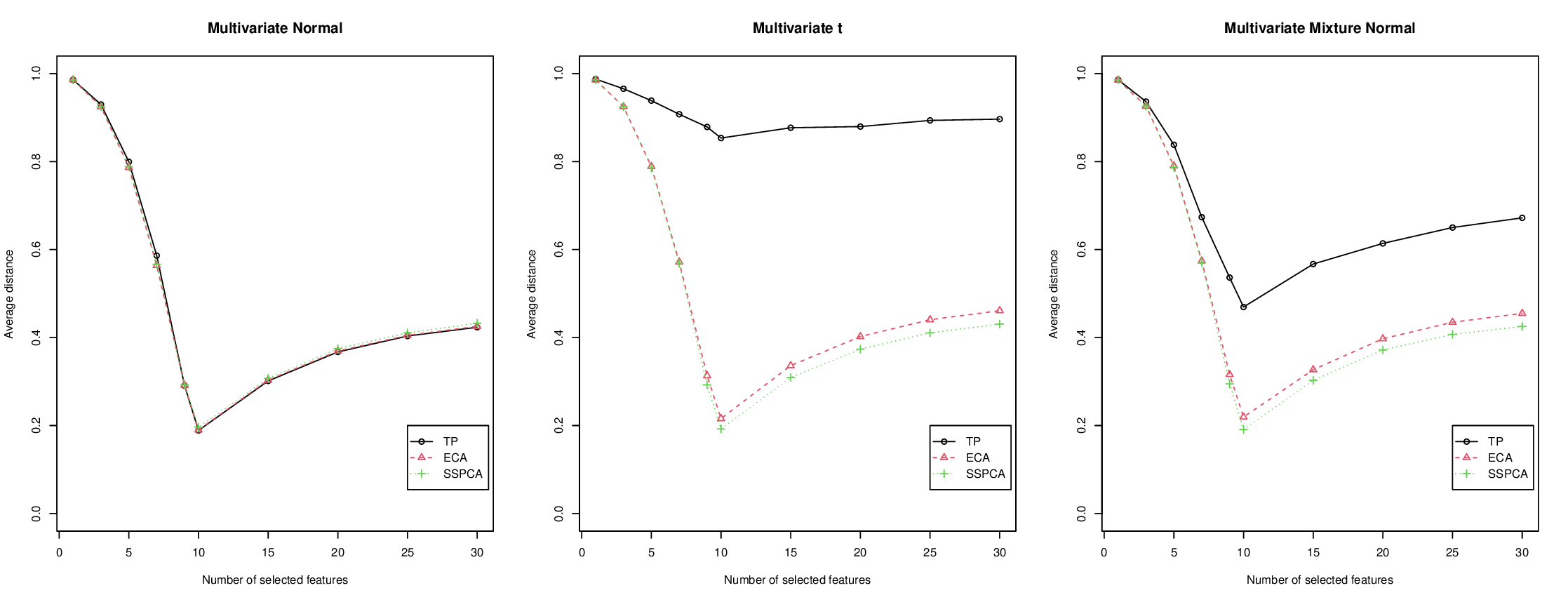}}
\end{figure}

Next, we consider estimating the top $m$ leading eigenvectors of the covariance matrix ${\bf \Sigma}$. Here we set $m=4$, the eigenvalues $\omega_1=10.1,\omega_2=6.2,\omega_3=3.3,\omega_4=1.4,\omega_5=\cdots=\omega_d=0.5$ and the cardinalities $s_1=s_2=10,s_3=s_4=8$. Figure \ref{figadm} plots the average distances $\frac{1}{4}\sum_{i=1}^4 |\sin\angle\left(\widehat{\boldsymbol{v}}_i, \boldsymbol{v}_i\right)|$
aginst the numbers of estimated nonzero entries $\frac{1}{4}\sum_{j=1}^4\left\|\widehat{\boldsymbol{v}}_j\right\|_0$ with $n=50,100,200$ and $d=100$. For simplicity, here we set $\left\|\widehat{\boldsymbol{v}}_j\right\|_0$ are all equal. Our findings indicate that as the sample size increases, the estimating errors become smaller, which aligns with the results presented in Figure \ref{figad2}. This observation suggests that the accuracy of our estimations improves with larger sample sizes. Furthermore, similar to the results observed in Figure \ref{figad1}, our proposed SSPCA method consistently outperforms the other two methods when dealing with heavy-tailed distributions. This indicates that our SSPCA method is particularly effective in handling data with heavy-tailed distributions, which are common in many real-world datasets.

In addition, under normal distribution, our proposed SSPCA method performs similarly to the TP method. This is an important finding because it suggests that our SSPCA method can achieve comparable performance to existing methods in standard scenarios, while also demonstrating superior performance in more challenging scenarios with heavy-tailed distributions.

\begin{figure}[htbp]
\centering
\caption{Histogram of estimator of the number of selected features with $n=400,d=100$. \label{figss2}}
\subfloat[Multivariate Normal]{\includegraphics[width=0.85\textwidth]{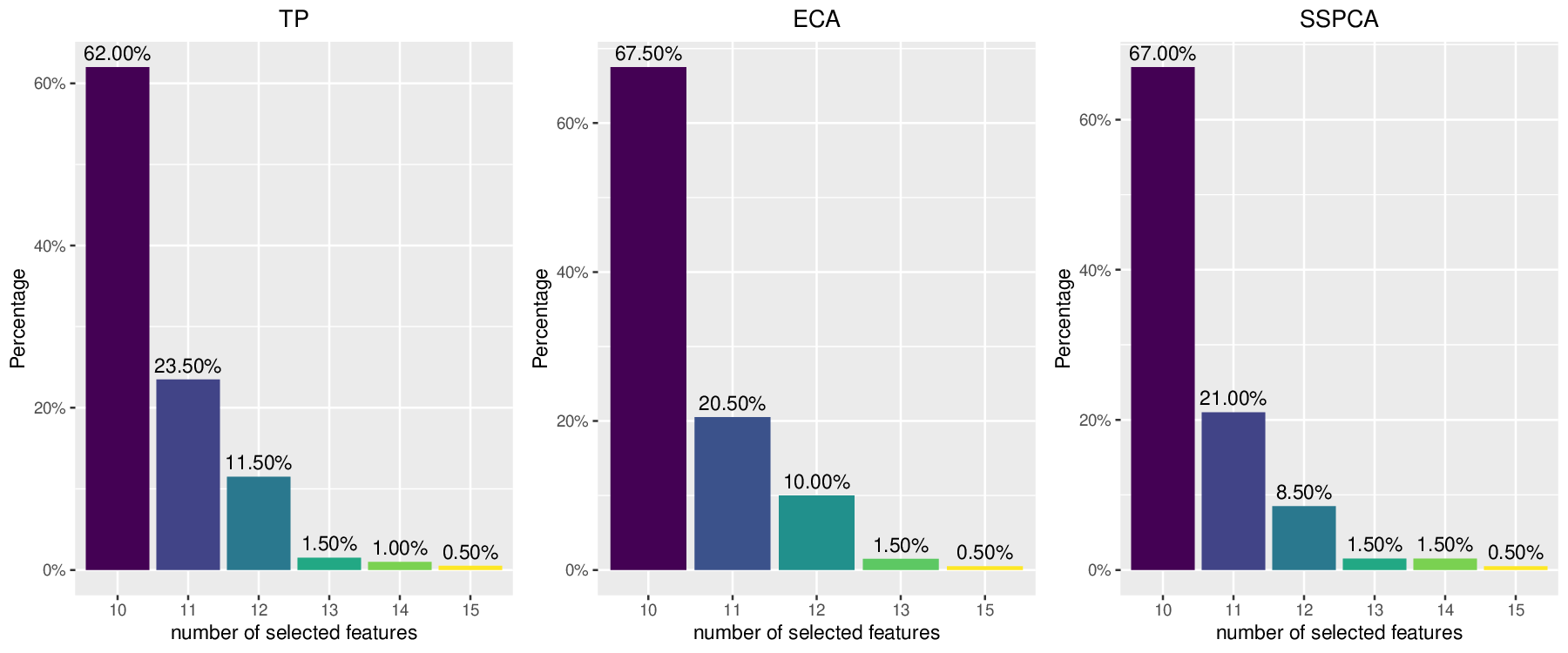}}\\
\subfloat[Multivariate t]{\includegraphics[width=0.85\textwidth]{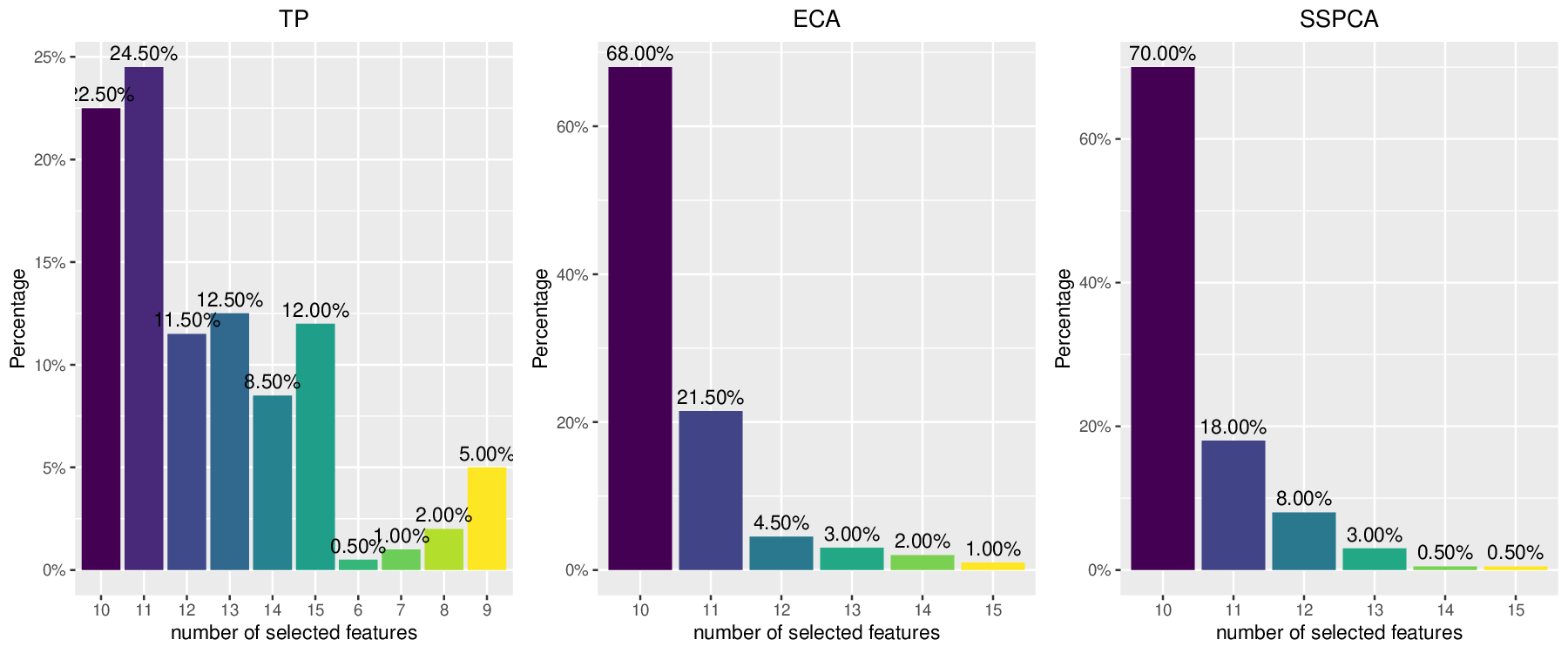}}\\
\subfloat[Multivariate Mixture Normal]{\includegraphics[width=0.85\textwidth]{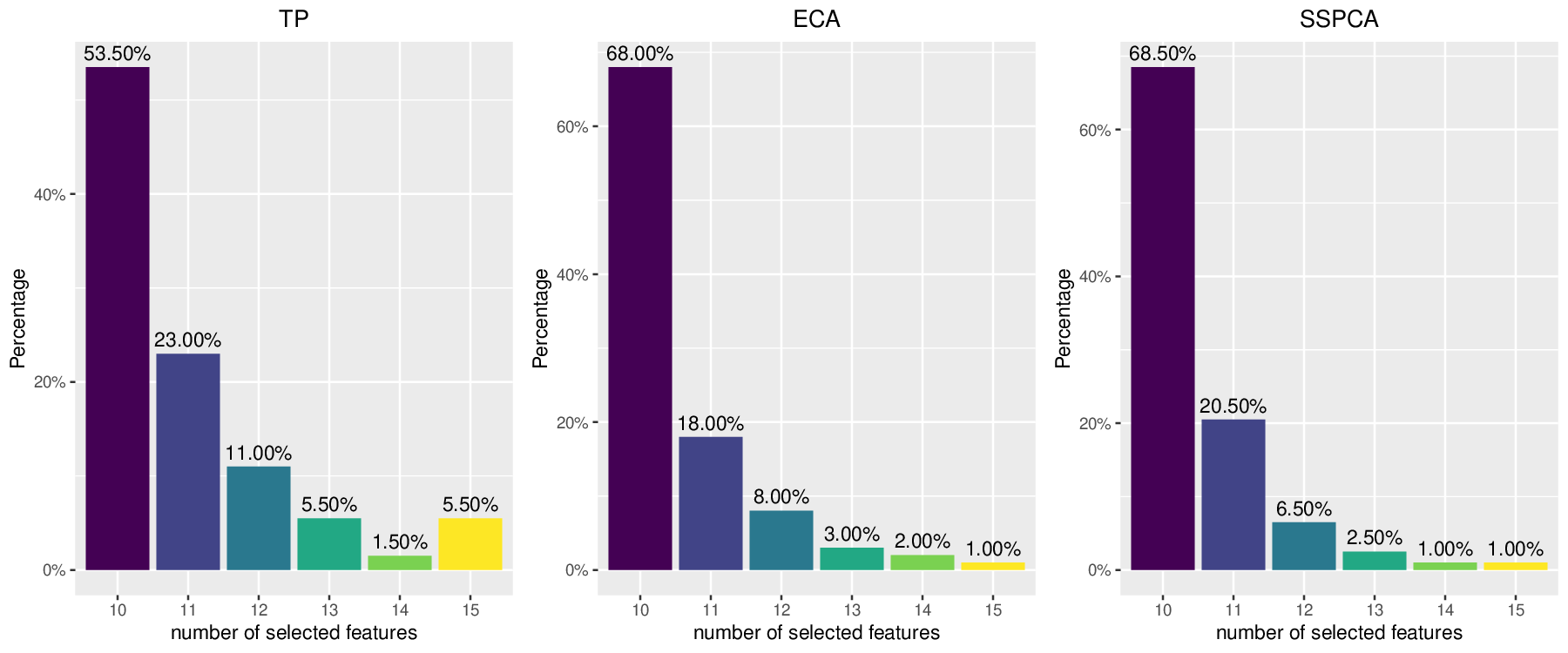}}
\end{figure}

Next, we further investigate the impact of estimating the number of selected features on the accuracy of the average distances. In this analysis, we select $\left\|\widehat{\boldsymbol{v}}_1\right\|_0$ as a representative and set all other values equal to it. Table \ref{tm2} reports the averaged distances between the estimates and true parameters of the top $m$ leading eigenvectors, comparing the estimated number of selected features $\hat{s}$ with the true number of selected features $s$. Figure \ref{figss2} presents the histogram of the estimated number of selected features with $n=400$ and $d=100$. The results obtained are consistent with those from the previous subsection. Specifically, the average distance with the estimated $\hat{s}$ is slightly larger than the oracle estimator with the true $s$.

\begin{table}[!ht]
\begin{center}
\caption{\label{tm2} The averaged distances between the estimates and true parameters of the top $m$ leading eigenvector with estimated number of selected features $\hat{s}$ and the true number of selected features $s$.}
                     \vspace{0.5cm}
                     \renewcommand{\arraystretch}{0.8}
                     \setlength{\tabcolsep}{3pt}{
\begin{tabular}{c|cc|cc|cc|cc|cc|cc}
\hline \hline
 $T$ & \multicolumn{6}{c}{{$n=200$}} & \multicolumn{6}{c}{{$n=400$}}\\ \hline
Distributions &\multicolumn{2}{c}{(I)}&\multicolumn{2}{c}{(II)}&\multicolumn{2}{c}{(III)} &\multicolumn{2}{c}{(I)}&\multicolumn{2}{c}{(II)}&\multicolumn{2}{c}{(III)}\\ \hline
 & $\hat{s}$&$s$& $\hat{s}$&$s$& $\hat{s}$&$s$& $\hat{s}$&$s$& $\hat{s}$&$s$& $\hat{s}$&$s$\\ \hline
 &\multicolumn{12}{c}{$d=100$}\\ \hline
TP   &0.184&0.168&0.859&0.847&0.483&0.446&0.128&0.115&0.790&0.767&0.268&0.237\\
ECA  &0.188&0.176&0.245&0.233&0.236&0.217&0.127&0.116&0.136&0.123&0.140&0.127\\
SSPCA&0.195&0.182&0.216&0.204&0.209&0.196&0.130&0.120&0.131&0.121&0.128&0.117\\
\hline
\hline
\end{tabular}}
\end{center}
\end{table}

Overall, these results demonstrate the robustness and effectiveness of our proposed SSPCA method in handling various types of data distributions. The ability to accurately estimate parameters even in the presence of heavy-tailed distribution is a valuable characteristic of our method, and it highlights its potential for use in a wide range of applications where data may not always follow a normal distribution.

\section{Real Data Analysis}
{ \subsection{S\&P 500 Index Stock Data}}
In this { subsection}, we apply three methodologies—Thresholding Pursuit (TP), Exponential Component Analysis (ECA), and Sparse and Structured Principal Component Analysis (SSPCA)—to analyze the Standard \& Poor's 500 (S\&P 500) index. To account for the dynamic nature of the index's composition over time, we compiled monthly returns for all securities included in the S\&P 500 from January 2005 to November 2018 ($n=165$). Given the evolving nature of the index, we focused on a consistent subset of $d = 374$ securities that were present throughout this entire period. As demonstrated in \cite{liu2023high}, stock returns exhibit non-Gaussian, heavy-tailed characteristics, which necessitate the use of robust statistical procedures. For simplicity, our analysis considers only the first two principal components.

\begin{figure}[ht]
	\centering
	\caption{Plots of principal components 1 (PC1) against principal components 2 (PC2) with three methods--TP, ECA and SSPCA. Here red dots represent the points with strong leverage influence. \label{figdata1}}
	\includegraphics[width=1\textwidth]{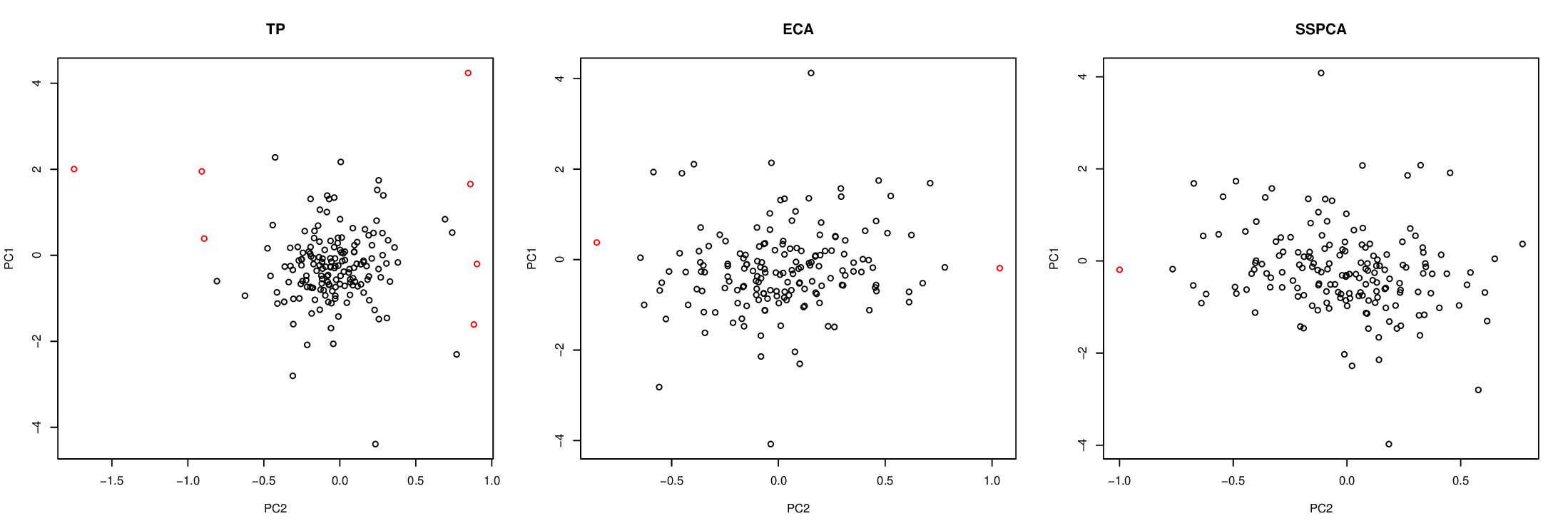}
\end{figure}

Utilizing a tuning parameter selection procedure, we determined optimal values of $k=d$ for the first principal component and $k=150$ for the second principal component. Figure \ref{figdata1} displays scatter plots of the first principal component (PC1) versus the second principal component (PC2) for each of the three methodologies—TP, ECA, and SSPCA.

Consistent with the approach in \cite{han2018eca}, red dots in the plots represent potential outliers with strong leverage influence. Leverage strength, defined as the diagonal values of the hat matrix in a linear regression model where the first principal component is regressed on the second, serves as an indicator of the impact of individual data points on the regression estimates \citep{neter1996}. High leverage strength implies that the inclusion of these points will significantly affect the linear regression estimates applied to the principal components of the data. We chose a threshold value of 0.05 to identify data points with strong leverage influence. Our analysis revealed that 6 data points have strong leverage influence for the TP method, 2 for the ECA method, and only 1 for the SSPCA method. These findings highlight the robustness of our proposed SSPCA method.

\begin{figure}[htbp]
	\centering
	\caption{Leverage influence of each method over time period. \label{figdata2}}
	\includegraphics[width=1\textwidth]{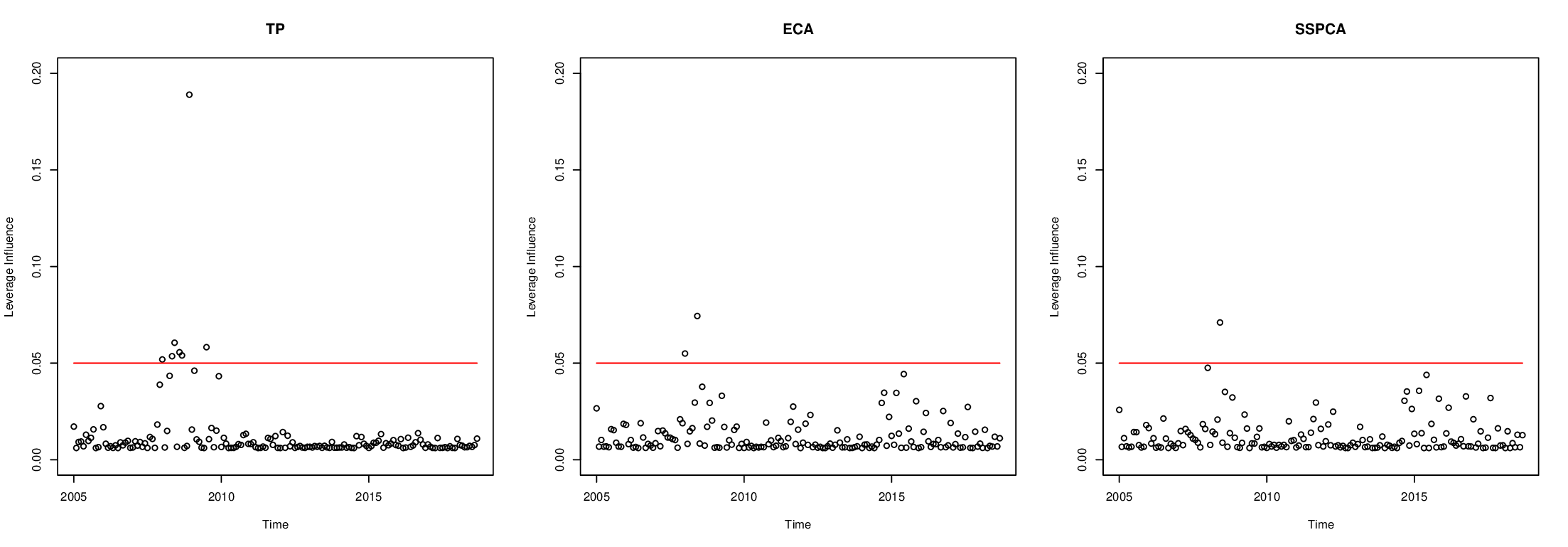}
\end{figure}

Furthermore, we examined the leverage influence of each data point over time for each methodology, as depicted in Figure \ref{figdata2}. We observed that data points with strong leverage influence tend to cluster around periods of financial crisis, indicating that these observations could have a profound impact on statistical inference. Notably, our SSPCA method exhibits reduced sensitivity to these outliers compared to the other methodologies. This robustness is particularly advantageous in financial applications, where outliers and extreme events are common and can significantly affect analysis results.

\subsection{MNIST Dataset}

{ In this subsection, we apply three methodologies to analyze the MNIST dataset \citep{lecun2002gradient}. The MNIST dataset comprises 60,000 grayscale images of handwritten digits ranging from zero to nine. Each image is $28 \times 28$ pixels in size and is labeled with its corresponding digit. For our analysis, we construct the training matrix using the first 660 samples of the digit “1” and the first 33 samples of the digit “7,” resulting in a $693 \times 784$ matrix. The remaining samples labeled as “1” and “7” constitute the test set, forming a $12,314 \times 784$ matrix. The training data are standardized (zero mean, unit variance), and the same parameters are used to standardize the test data.

For simplicity, we focus on the first two principal components in each method. Using a parameter selection procedure over the candidate set $k \in \{300, 350, 400, 450, 500, 550, 600\}$, we determine the optimal values for the first and second principal components for each of the three methods. Subsequently, the two principal components derived from each method are used to train a Support Vector Machine (SVM) classifier on the training set. The trained model is then evaluated on the test set, yielding the following classification accuracies: 0.4939 for TP, 0.8221 for ECA, and 0.8589 for SSPCA. Figure \ref{figdata} presents the scatter plots of the first (PC1) and second (PC2) principal components obtained from TP, ECA, and SSPCA, respectively.  In each plot, red dots indicate potential outliers with strong leverage effects. A threshold of 0.02 is employed to identify such influential data points. Our analysis revealed that 14 data points have strong
leverage influence for the TP method, 3 for the ECA method, and 0 for the SSPCA method. These results demonstrate the robustness of SSPCA in mitigating the influence of outliers.

 Additionally, we conduct 100 simulation experiments.  In each experiment, 660 samples labeled “1” and 33 samples labeled “7” are randomly drawn from the training set to form the training data, with the remaining “1” and “7” samples used for testing.  The same standardization and parameter selection procedure as described above is applied in each experiment.  We compute the average classification accuracy across the 100 runs, yielding the following results: 0.4939 for TP, 0.7154 for ECA, and 0.8074 for SSPCA.

These results demonstrate the advantages of the proposed SSPCA method. It consistently outperforms TP and ECA in classification accuracy and shows greater robustness to outliers, as evidenced by the absence of high-leverage points. This indicates that SSPCA provides more stable and reliable representations in high-dimensional settings.

\begin{figure}[ht]
\centering
\caption{Plots of principal components 1 (PC1) against principal components 2 (PC2) with three methods--TP, ECA and SSPCA. Here red dots represent the points with strong leverage influence. \label{figdata}}
\includegraphics[width=1\textwidth]{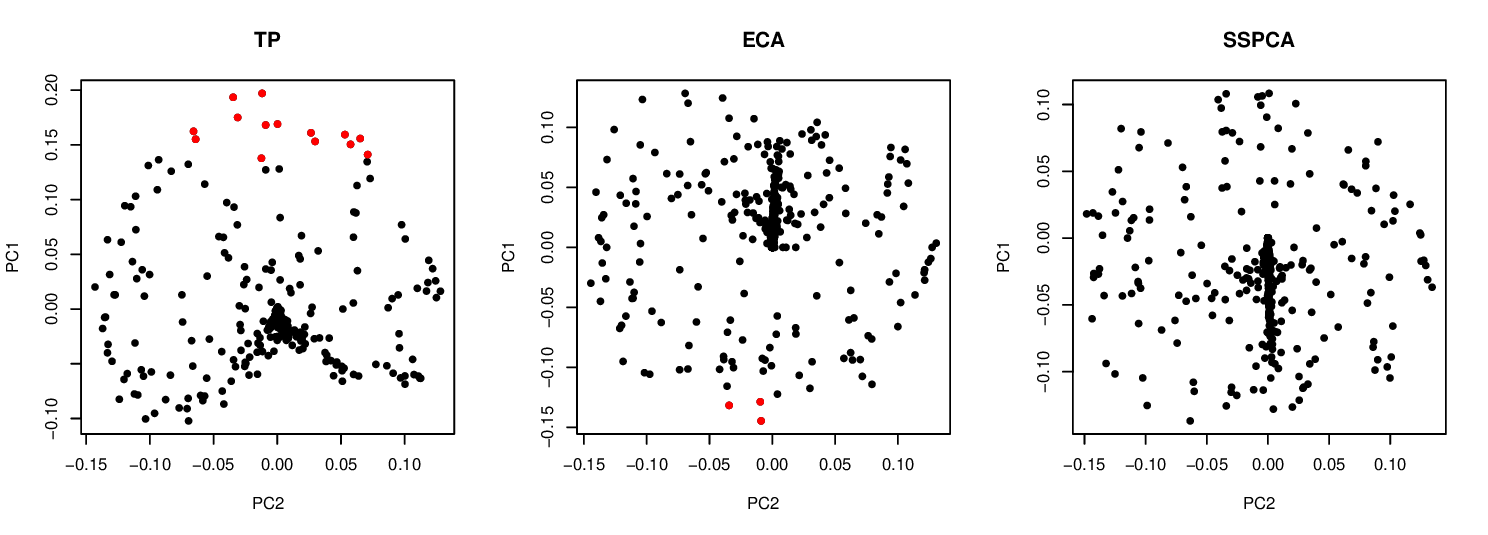}
\end{figure}

}

\section{Conclusion}
In this paper, we analyze the application of principal component analysis (PCA) with a sample spatial-sign covariance matrix in high-dimensional contexts. We determine the approximation errors of the principal component estimator under both non-sparse and sparse conditions. Simulation studies and real-world data applications demonstrate the computational efficiency and robustness of our proposed methods. PCA is a widely utilized technique in numerous fields, and therefore, the methods presented in this paper can be applied to various applications, including high-dimensional factor analysis \citep{he2022large} and dimension reduction \citep{chen2022high}.

\newpage

\setcounter{section}{0}

\renewcommand\thefigure{S\arabic{figure}}
\renewcommand\thetable{S\arabic{table}}
\renewcommand\thesection{S\arabic{section}}

{\Large Supplemental Material of ``Spatial Sign based Principal Component Analysis for High Dimensional Data"}

\section{Appendix A: Fantope Projection}
Similar to \cite{VuChoLeiRohe2013}, we define $\Y_1$ as the solution to the following
convex program:
$$
\begin{aligned}
\bm {Y}_1:= & \underset{\mathbf{M} \in \mathbb{R}^{d \times d}}{\arg \max }\langle\widehat{\mathbf{S}}, \mathbf{M}\rangle-\lambda \sum_{j, k}\left|\mathbf{M}_{j k}\right|, \text { subject to } \mathbf{0} \preceq \mathbf{M} \preceq \mathbf{I}_d \text { and } \operatorname{Tr}(\mathbf{M})=1,
\end{aligned}
$$
where for any two matrices $\mathbf{A}, \mathbf{B} \in \mathbb{R}^{d \times d}, \mathbf{A} \preceq \mathbf{B}$ represents that $\mathbf{B}-\mathbf{A}$ is positive semidefinite. Here, $\{\mathbf{M}: \mathbf{0} \preceq \mathbf{M} \preceq$ $\left.\mathbf{I}_d, \operatorname{Tr}(\mathbf{M})=1\right\}$ is a convex set called the Fantope.  The initial parameter $\boldsymbol{v}^{(0)}$ then, is the normalized vector consisting of the largest entries in $\boldsymbol{u}_1\left(\mathbf{Y}_1\right)$, where $\mathbf{Y}_1$ is calculated in (5.1):
\begin{align}\label{54}
\boldsymbol{v}^{(0)} & =\boldsymbol{w}^0 /\left\|\boldsymbol{w}^0\right\|_2, \text { where } \boldsymbol{w}^0=\operatorname{TRC}\left(\boldsymbol{u}_1\left(\mathbf{Y}_1\right), J_\varphi\right) \text { and }
J_\varphi & =\left\{j:\left|\left(\boldsymbol{u}_1\left(\mathbf{Y}_1\right)\right)_j\right| \geq \varphi\right\}
\end{align}
We have $\left\|\boldsymbol{v}^{(0)}\right\|_0=\operatorname{supp}\left\{j:\left|\left(\boldsymbol{u}_1\left(\mathbf{Y}_1\right)\right)_j\right|>0\right\}$.
To show the consistency of the initial estimator, we need the following assumptions:
\begin{itemize}
\item[(C1)] $\operatorname{rank}(\boldsymbol{\Sigma})=q$ and $\left\|\boldsymbol{u}_1(\boldsymbol{\Sigma})\right\|_0 \leq s$. Additionally, we assume $\nu_i$ is sub-gaussian distributed, i.e. $\|\nu_i\|_{\psi_2}\le K_\nu<\infty$.
\item[(C2)] $\lambda_1(\boldsymbol{\Sigma}) / q \lambda_q(\boldsymbol{\Sigma})=O\left(\lambda_1(\mathbf{K})\right),\|\boldsymbol{\Sigma}\|_{\mathrm{F}} \log d=o(\operatorname{Tr}(\boldsymbol{\Sigma}))$. $\lambda_2(\boldsymbol{\Sigma}) / \lambda_1(\boldsymbol{\Sigma})$ is upper bounded by an absolute constant less than 1 , and $\lambda \asymp \lambda_1(\mathbf{K}) \sqrt{\log d / n}$.
\item[(C3)] let $J_0:=\left\{j:\left|\left(\boldsymbol{u}_1(\mathbf{K})\right)_j\right|=\Omega^0(s \log d / \sqrt{n})\right\}$. Set $\varphi$ in (\ref{54}) to be $\varphi=C_2 s(\log d) / \sqrt{n}$ for some positive absolute constant $C_2$. If $s \sqrt{\log d / n} \rightarrow 0$, and $\left\|\left(\boldsymbol{u}_1(\mathbf{K})\right)_{J_0}\right\|_2 \geq C_3>0$ is lower bounded by an absolute positive constant.
\end{itemize}

\begin{theorem}\label{thad}
Under Assumption (A1)-(A2) and (C1), if $(\log d+\log (1 / \alpha)) / n \rightarrow 0$ and $\lambda \geq C_1\left(\frac{8 \lambda_1(\boldsymbol{\Sigma})}{q \lambda_q(\boldsymbol{\Sigma})}+\|\mathbf{S}\|_{\max }\right) \sqrt{\frac{\log d+\log (1 / \alpha)}{n}}$, for sufficient large $n$, we have
$$
\left|\sin \angle\left(\boldsymbol{u}_1\left(\mathbf{Y}_1\right), \boldsymbol{u}_1(\mathbf{S})\right)\right| \leq \frac{8 \sqrt{2} s \lambda}{\lambda_1(\mathbf{S})-\lambda_2(\mathbf{K})},
$$
with probability larger than $1-\alpha^2$. Additionally, if condition (C2) also hold, we have
$$
\left|\sin \angle\left(\boldsymbol{u}_1\left(\mathbf{Y}_1\right), \boldsymbol{u}_1(\mathbf{S})\right)\right|=O_p\left(s \sqrt{\frac{\log d}{n}}\right).
$$
Then, if condition (C3) also hold,  with probability tending to $1,\left\|\boldsymbol{v}^{(0)}\right\|_0 \leq s$ and $\left|\left(\boldsymbol{v}^{(0)}\right)^T \boldsymbol{u}_1(\mathbf{S})\right|$ is lower bounded by $C_3/2$.
\end{theorem}

\begin{figure}[htbp]
\centering
\caption{Curves of averaged distances between the estimates and true parameters with different number of selected features on estimating leading eigenvectors. \label{figfp}}
\subfloat[$n=200,d=100$]{\includegraphics[width=0.85\textwidth]{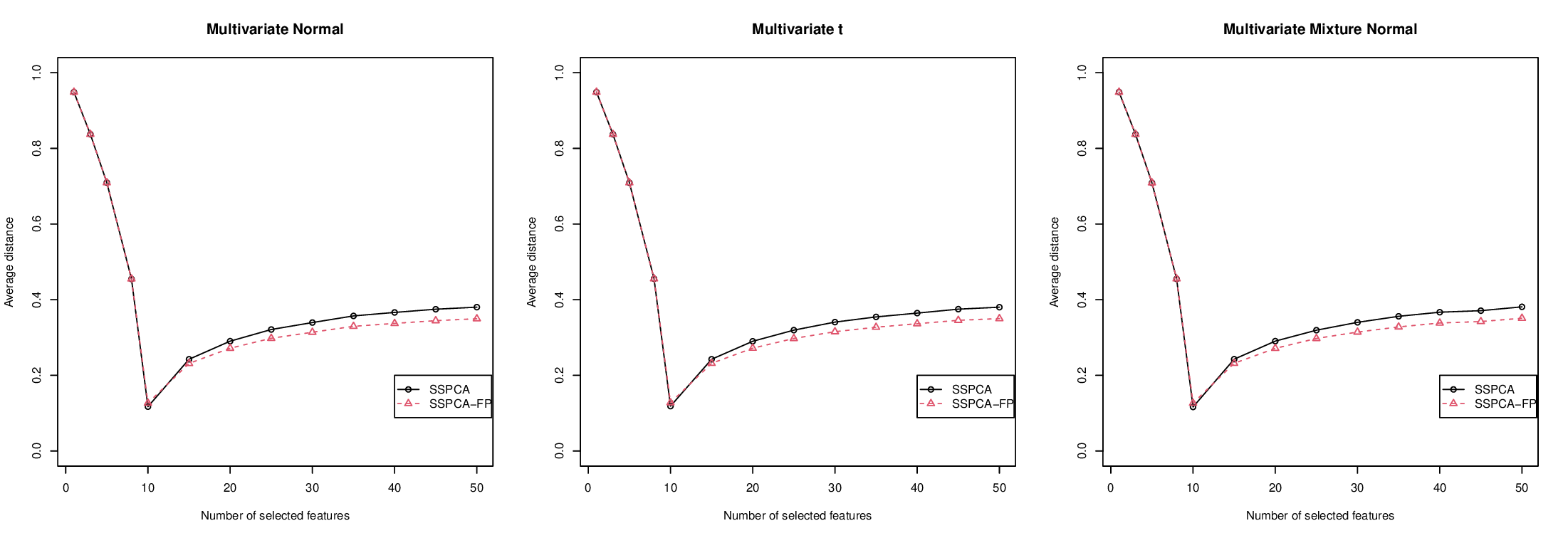}}\\
\subfloat[$n=200,d=200$]{\includegraphics[width=0.85\textwidth]{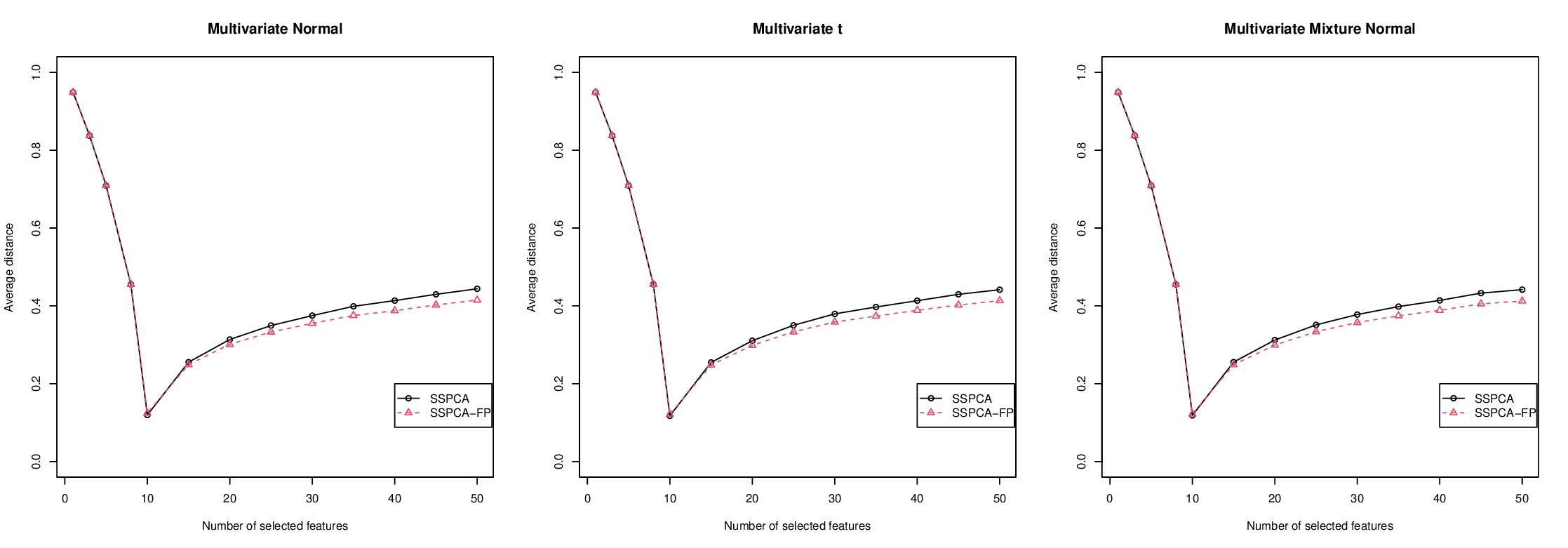}}
\end{figure}

Next, we conduct simulation studies to compare the two distinct initial estimators. SSPCA denotes the proposed method utilizing the eigenvector as the initial estimator, whereas SSPCA-FP refers to the proposed method incorporating Fantope Projection. In this analysis, we focus solely on the estimation of the first eigenvector. Figure \ref{figfp} presents the outcomes of these two approaches. Our observations reveal that when the number of selected features matches the true number, SSPCA-FP performs similar to SSPCA. However, when the number of selected features exceeds the true parameters, SSPCA-FP exhibits smaller averaged distances compared to SSPCA. We also compare these two methods with the averaged distances between the estimates and true parameters of the leading eigenvector with estimated number of selected features $\hat{s}$. Table \ref{s1} reveals the simulation results with the same settings as Table \ref{t2}. Overall, the performance of these two initial estimators is quite comparable. SSPCA-FP exhibits slightly smaller averaged distances when compared to the estimator $\hat s$. Therefore, if computational time is not a constraint, we recommend using SSPCA-FP as the primary choice.

\begin{table}[!ht]
\begin{center}
\caption{\label{s1} The averaged distances between the estimates and true parameters of the leading eigenvector with estimated number of selected features $\hat{s}$ and the true number of selected features $s$.}
                     \vspace{0.5cm}
                     \renewcommand{\arraystretch}{0.8}
                     \setlength{\tabcolsep}{3pt}{
\begin{tabular}{c|cc|cc|cc|cc|cc|cc}
\hline \hline
 $T$ & \multicolumn{6}{c}{{$n=200$}} & \multicolumn{6}{c}{{$n=400$}}\\ \hline
Distributions &\multicolumn{2}{c}{(I)}&\multicolumn{2}{c}{(II)}&\multicolumn{2}{c}{(III)} &\multicolumn{2}{c}{(I)}&\multicolumn{2}{c}{(II)}&\multicolumn{2}{c}{(III)}\\ \hline
 & $\hat{s}$&$s$& $\hat{s}$&$s$& $\hat{s}$&$s$& $\hat{s}$&$s$& $\hat{s}$&$s$& $\hat{s}$&$s$\\ \hline
 &\multicolumn{12}{c}{$d=100$}\\ \hline
SSPCA   &0.141&0.121&0.140&0.119&0.148&0.125&0.091&0.083&0.090&0.082&0.094&0.085\\
SSPCA-FP&0.140&0.125&0.143&0.129&0.145&0.126&0.090&0.085&0.091&0.081&0.093&0.085\\ \hline
 &\multicolumn{12}{c}{$d=200$}\\ \hline
SSPCA   &0.141&0.115&0.139&0.117&0.141&0.116&0.087&0.081&0.088&0.080&0.086&0.081\\
SSPCA-FP&0.142&0.116&0.138&0.117&0.140&0.119&0.086&0.083&0.087&0.082&0.086&0.082\\ \hline
 &\multicolumn{12}{c}{$d=300$}\\ \hline
SSPCA   &0.142&0.118&0.145&0.118&0.150&0.120&0.087&0.081&0.089&0.084&0.088&0.083\\
SSPCA-FP&0.141&0.120&0.139&0.120&0.141&0.117&0.086&0.083&0.087&0.085&0.086&0.083\\
\hline
\hline
\end{tabular}}
\end{center}
\end{table}

\section{Appendix B: Proof of Theorems}
\subsection{Some useful lemmas}
The accuracies of constant and linear approximations of function $|\mathbf{y}-\mu|^{-1}(\mathbf{y}-\mu)$ of $\mu$ are given by \cite{Oja2010Multivariate}.
\begin{lemma}\label{le1}
Let $\mathbf{y} \neq \mathbf{0}$ and $\mu$ be any $p$-vectors, $p>1$. Write also
$
r=|\mathbf{y}| \text { and } \mathbf{u}=|\mathbf{y}|^{-1} \mathbf{y} .
$
\begin{align*}
&\left|\frac{\mathbf{y}-\mu}{|\mathbf{y}-\mu|}-\frac{\mathbf{y}}{|\mathbf{y}|}\right| \leq 2 \frac{|\mu|}{r}\\
&\left|\frac{\mathbf{y}-\mu}{|\mathbf{y}-\mu|}-\frac{\mathbf{y}}{|\mathbf{y}|}-\frac{1}{r}\left[\mathbf{I}_p-\mathbf{u u ^ { \prime }}\right] \mu\right| \leq C \frac{|\mu|^{1+\delta}}{r^{1+\delta}}
\end{align*}
for all $0<\delta<1$ where $C$ does not depend on $\mathbf{y}$ or $\mu$.
\end{lemma}

The following lemma is the result of the Lemma 19 of \cite{arcones1998asymptotic}. See also \cite{bai1990asymptotic} and \cite{Oja2010Multivariate}.
\begin{lemma}\label{le2}
The accuracies of constant, linear and quadratic approximations of the function $\boldsymbol{\mu} \mapsto\|\mathbf{z}-\boldsymbol{\mu}\|{ _2}$ can be given by
\begin{itemize}
\item[(1)] $|\|\mathbf{z}-\boldsymbol{\mu}\|{ _2}-\|\mathbf{z}\|{ _2}| \leq\|\boldsymbol{\mu}\|{ _2}$,
\item[(2)] $\left|\|\mathbf{z}-\boldsymbol{\mu}\|{ _2}-\|\mathbf{z}\|{ _2}+\mathbf{u}^{T} \boldsymbol{\mu}\right| \leq 2 r^{-1}\|\boldsymbol{\mu}\|{ _2}^2$,
\item[(3)] $\left|\|\mathbf{z}-\boldsymbol{\mu}\|{ _2}-\|\mathbf{z}\|{ _2}+\mathbf{u}^{T} \boldsymbol{\mu}-\boldsymbol{\mu}^{T}(2 r)^{-1}\left[\mathbf{I}_p-\mathbf{u} \mathbf{u}^{T}\right] \boldsymbol{\mu}\right| \leq c r^{-1-\delta}\|\boldsymbol{\mu}\|{ _2}^{2+\delta}$
for all $0<\delta<1$,
\end{itemize}
where $\mathbf{z}=r \mathbf{u}, r=\|\mathbf{z}\|{ _2}, \mathbf{u}=\|\mathbf{z}\|{ _2}^{-1} \mathbf{z}$ and the constant $c$ does not depend on $\mathbf{z}$ or $\boldsymbol{\mu}$.
\end{lemma}

Next, we restate Theorem 1.4 in \cite{tropp2012}.
\begin{lemma}\label{mb} (Matrix Bernstein) Consider a finite sequence $\left\{\X_k\right\}$ of independent, random, self-adjoint matrices with dimension $d$. Assume that each random matrix satisfies
$$
E \X_k=\mathbf{0} \quad \text { and } \quad \lambda_{\max }\left(\X_k\right) \leq R \quad \text { almost surely }
$$
Then, for all $t \geq 0$,
$$
P\left\{\left\|\sum_k \X_k\right\|_2 \geq t\right\} \leq d \cdot \exp \left(\frac{-t^2 / 2}{\sigma^2+R t / 3}\right) \quad \text { where } \sigma^2:=\left\|\sum_k E\left(\X_k^2\right)\right\|_2.
$$
\end{lemma}

\begin{lemma}\label{mu}
Under conditions (A1)-(A2), we have $\hat{\bmu}-\bmu=O_p(\zeta_1^{-1}n^{-1/2})$ and
\begin{align*}
\hat{\bmu}-\bmu=\zeta_1^{-1}\frac{1}{n}\sum_{i=1}^n \U_i+o_p(\zeta_1^{-1}n^{-1/2}).
\end{align*}
\end{lemma}

{\bf Proof:}
Define the object function
\begin{align*}
L(\bth)=\sum_{i=1}^n\|\X_i-\bmu-\bth\|{ _2}-\sum_{i=1}^n\|\X_i-\bmu\|{ _2}
\end{align*}
and $\hbth=\arg\min L(\bth)$. Then $\hbth=\hat{\bmu}-\bmu$. Next, we proof that for any $\epsilon>0$, there exists $C>0$ such that
\begin{align*}
\lim \inf_n P\left(\inf_{\u \in \mathbb{S}^{d-1}}L(C\zeta_1^{-1}n^{-1/2}\u)>0\right)\ge 1-\epsilon
\end{align*}
for large enough $n$. Then, by the convexity of $L(\cdot)$, we can obtain $$P\left(\|\hbth\|{ _2}\le C \zeta_1^{-1}n^{-1/2}\right)\ge 1-\epsilon,$$
which means $\hbth=O_p(\zeta_1^{-1}n^{-1/2})$.
 By Lemma \ref{le2}, we have
\begin{align*}
L(C\zeta_1^{-1}n^{-1/2}\u) \ge &-{C}\sum_{i=1}^n (\zeta_1^{-1}n^{-1/2}\U_i^T \u)\\
&+{C^2}\sum_{i=1}^n \zeta_1^{-2}n^{-1} \frac{1}{2r_i}\u^T[\I_d-\U_i\U_i^T]\u\\
&-\sum_{i=1}^n cC^{2+\delta}\frac{1}{r_i^{1+\delta}} \zeta_1^{-(2+\delta)}n^{-1-\frac{1}{2}\delta}\\
&\doteq A_1+A_2+A_3.
\end{align*}
Because $(\U_i^T \u)^2\le 1$, $E(\U_i^T \u)=0$ and $\var(\U_i^T\u)=\u^T\S\u\le \|\S\|_2\le \tr(\S)=1$, so, by Chebyshev inequality, for sufficient large $M>0$, we have
\begin{align*}
P\left(\sum_{i=1}^n n^{-1/2}\U_i^T \u>M\right)\le \frac{1}{M}\le\frac{\epsilon}{3}.
\end{align*}
Thus, with at least probability $1-\frac{\epsilon}{3}$, we have
\begin{align*}
-{C}\sum_{i=1}^n (\zeta_1^{-1}n^{-1/2}\U_i^T \u) \ge -C \zeta_1^{-1}M.
\end{align*}
In addition, $E(1-(\U_i^T\u)^2)=1-\u^T\S\u \ge 1-\|\S\|_2>\varphi>0$ by condition (A2), $\var(\zeta_1^{-1}r_i^{-1}[1-(\U_i^T\u)^2])\le E(\zeta_1^{-2}r_i^{-2}) E((\U_i\u)^4)-(\u^T\S\u)^2\le \zeta_1^{-2}\zeta_2 \le \zeta$. Thus, by Chebyshev inequality, we have
\begin{align*}
&P\left(\frac{1}{n}\sum_{i=1}^n \zeta_1^{-1} \frac{1-(\U_i^T\u)^2}{2r_i}-\frac{1}{2}(1-\u^T\S\u)\le -\frac{1}{4}(1-\u^T\S\u) \right)\\
& \le \frac{16\var(\zeta_1^{-1}r_i^{-1}[1-(\U_i^T\u)^2])}{n(1-\u^T\S\u)^2}\le \frac{16\zeta}{n\psi^2}\le \frac{\epsilon}{3}
\end{align*}
for sufficient large $n$.
Thus, with at least probability $1-\frac{\epsilon}{3}$, we have
\begin{align*}
A_2\ge \frac{C^2}{4\zeta_1}(1-\u^T\S\u)\ge \frac{C^2\psi}{4\zeta_1}.
\end{align*}
Finally, by the Chebyshev inequality, we have
\begin{align*}
P\left(\frac{1}{n}\sum_{i=1}^n\frac{1}{r_i^{1+\delta}} \zeta_1^{-(1+\delta)} \ge \frac{1}{2}E(\nu_i^{1+\delta}) \right) \le \frac{4\var(\nu_i^{1+\delta})}{n[E(\nu_i^{1+\delta})]^2}\le \frac{4\zeta}{n}\le \frac{\epsilon}{3}
\end{align*}
by condition (A1) for sufficient large $n$. So $A_3\ge -\frac{cC^{2+\delta}}{2\zeta_1}E(\nu_i^{1+\delta})n^{-\frac{1}{2}\delta}$.
Thus, at least probability $1-\epsilon$, we have
\begin{align*}
\zeta_1 L(\zeta_1^{-1}n^{-1/2+\epsilon}\u) \ge & -CM+\frac{C^2\psi}{4}-\frac{cC^{2+\delta}}{2}E(\nu_i^{1+\delta})n^{-\frac{1}{2}\delta}>0
\end{align*}
for large enough $n$ and $C$.

Next, we consider the equation $\sum_{i=1}^n U(\X_i-\hat{\bmu})=0$. Note that
\begin{align*}
\sum_{i=1}^n U(\X_i-\hat{\bmu})=\sum_{i=1}^n \frac{\X_i-\bmu-\hbth}{\|\X_i-\bmu-\hbth\|{ _2}}=\sum_{i=1}^n \frac{\U_i-r_i^{-1}\hbth}{(1-2r_i^{-1}\U_i^T\hbth+r_i^{-2}\|\hbth\|_2^2)^{1/2}}
\end{align*}
which implies that
{ 
	\begin{align*}
		&n^{-1} \sum_{i=1}^n\left(\U_i-r_i^{-1} \hat{\boldsymbol{\theta}}\right)\left(1-2 r_i^{-1} \U_i^{\top} \hat{\boldsymbol{\theta}}+r_i^{-2}\|\hat{\boldsymbol{\theta}}\|_2^2\right)^{-1 / 2}=0.
	\end{align*}
By the Taylor expansion, the above equation can be rewritten as
\begin{align*}
	&n^{-1} \sum_{i=1}^n\left(\U_i-r_i^{-1} \hat{\boldsymbol{\theta}}\right)\left(1+r_i^{-1} \U_i^{\top} \hat{\boldsymbol{\theta}}-2^{-1} r_i^{-2}\|\hat{\boldsymbol{\theta}}\|_2^2+\delta_{1 i}\right)=0
\end{align*}
where $\delta_{1 i}=O_p\big\{\big(r_i^{-1} \U_i^{\top} \hat{\boldsymbol{\theta}}-2^{-1} r_i^{-2}\|\hat{\boldsymbol{\theta}}\|_2^2\big)^2\big\}=O_p\left(n^{-1}\right)$. Then, we have
\begin{align*}
	& n^{-1} \sum_{i=1}^n\left(1-2^{-1} r_i^{-2}\|\hat{\boldsymbol{\theta}}\|_2^2+\delta_{1 i}\right) \U_i+n^{-1} \sum_{i=1}^n r_i^{-1}\left(\U_i^{\top}\hat{\boldsymbol{\theta}}\right) \U_i \\
	= & n^{-1} \sum_{i=1}^n r_i^{-1}\left(1-2^{-1} r_i^{-2}\|\hat{\boldsymbol{\theta}}\|_2^2+\delta_{1 i}\right)\hat{\boldsymbol{\theta}}+n^{-1} \sum_{i=1}^n r_i^{-2}\left(\U_i^{\top}\hat{\boldsymbol{\theta}}\right)\hat{\boldsymbol{\theta}},
\end{align*}
which implies
\begin{align*}
	&n^{-1} \sum_{i=1}^n\left(1-2^{-1} r_i^{-2}\|\hat{\boldsymbol{\theta}}\|_2^2+\delta_{1 i}\right) \U_i+n^{-1} \sum_{i=1}^n r_i^{-1}\left(\U_i^{\top}\hat{\boldsymbol{\theta}}\right) \U_i
	= & n^{-1} \sum_{i=1}^n r_i^{-1}\left(1+\delta_{1 i}+\delta_{2 i}\right) \hat{\boldsymbol{\theta}},
\end{align*}
where $\delta_{2 i}=r_i^{-1} \U_i^{\top} \hat{\boldsymbol{\theta}}-2^{-1} r_i^{-2}\|\hat{\boldsymbol{\theta}}\|_2^2=O_p(\delta_{1 i}^{1 / 2})$. 
Then, we obtained that
$$
\hbth=\left\{\zeta_1+O_p\left(\zeta_1 n^{-1 / 2}\right)\right\}^{-1}\Big(n^{-1} \sum_{i=1}^{n} \mathbf{U}_i+\varrho\Big)=n^{-1} \zeta_1^{-1} \sum_{i=1}^{n} \mathbf{U}_i+O\left(\zeta_1^{-1}\varrho\right),
$$
 where $\|\varrho\|_2=O_p(n^{-1})$.}
\subsection{Proof of Theorem \ref{th1}}
Define $\tilde \S=\frac{1}{n}\sum_{i=1}^n \U_i\U_i^T$, $\U_i=U(\bm X_i-\bm \mu)$.
Obviously,
\begin{align*}
\left\|\U_i\U_i^T-\S\right\|_2\le \left\|\U_i\U_i^T\right\|_2+ \left\|\S\right\|_2= 1+\left\|\S\right\|_2,
\end{align*}
and
\begin{align*}
\left\|E[(\U_i\U_i^T-\S)^2]\right\|_2=\left\|\S-\S^2\right\|_2\le \left\|\S\right\|_2+\left\|\S\right\|_2^2.
\end{align*}
Thus, according to Lemma \ref{mb}, we have
\begin{align*}
P\left(\left\|\tilde \S-\S\right\|_2\ge t\right) &\le d \cdot \exp \left(\frac{-nt^2 / 2}{(\left\|\S\right\|_2+\left\|\S\right\|_2^2)+(1+\left\|\S\right\|_2) t / 3}\right)\\
&\le d \cdot \exp \left(\frac{-nt^2 }{4(\left\|\S\right\|_2+\left\|\S\right\|_2^2)}\right)
\end{align*}
for small enough $t\le 3\left\|\S\right\|_2$. Setting
$$
t=\sqrt{\frac{4\left(\|\mathbf{S}\|_2+\|\mathbf{S}\|_2^2\right)(\log d+\log (3 / \alpha))}{n}}=\|\mathbf{S}\|_2 \sqrt{\frac{4\left(1+r^*(\mathbf{S})\right)(\log d+\log (3    / \alpha))}{n}}
$$
we have $P\left(\left\|\tilde \S-\S\right\|_2\ge t\right) \le \alpha/3$.

Next, we will show the bound of $\|\hat{\S}-\tilde{\S}\|_2$. Define $\hat{\U}_i=U(\bm X_i-\hat{\bm \mu})$.
\begin{align*}
\hat{\S}-\tilde{\S}=&\frac{1}{n}\sum_{i=1}^n [\hat{\U}_i\hat{\U}_i^T-\U_i\U_i^T]\\
=&\frac{2}{n}\sum_{i=1}^n [\hat{\U}_i-\U_i]\U_i^T+\frac{1}{n}\sum_{i=1}^n [\hat{\U}_i-\U_i][\hat{\U}_i-\U_i]^T
\end{align*}
For any $\bm v\in \mathbb{S}^{d-1}$,
\begin{align*}
&\frac{1}{n}\sum_{i=1}^n \bm v^T[\hat{\U}_i-\U_i][\hat{\U}_i-\U_i]^T\bm v=\frac{1}{n}\sum_{i=1}^n([\hat{\U}_i-\U_i]^T\bm v)^2\\
&\le\frac{1}{n}\sum_{i=1}^n \left\|\hat{\U}_i-\U_i\right\|^2_2
\end{align*}
By Lemma \ref{le1}, we have
\begin{align*}
\left\|U(\bm X_i-\hat{\bm \mu})-U(\bm X_i-\bm \mu)\right\|_2\le 2\frac{\left\|\hat{\bm \mu}-\bm \mu\right\|_2}{r_i}.
\end{align*}
Thus,
\begin{align*}
\frac{1}{n}\sum_{i=1}^n \left\|\hat{\U}_i-\U_i\right\|^2_2 \le \frac{1}{n}\sum_{i=1}^n  r_i^{-2}\left\|\hat{\bm \mu}-\bm \mu\right\|_2^2.
\end{align*}
By lemma \ref{mu}, we have, there exist a positive constant $C_\alpha$ such that $\left\|\hat{\bm \mu}-\bm \mu\right\|_2^2\le C_\alpha^2\zeta_1^{-2}n^{-1}$ with probability larger than $1-\alpha/6$. Additionally, by Chebyshev inequality,
\begin{align*}
P\left(\frac{1}{n}\sum_{i=1}^n  r_i^{-2}\zeta_1^{-2} \ge 2E(\nu_i^2)\right)\le \frac{\kappa_\nu}{n (1+\sigma_\nu^2)^2}\le \frac{\alpha}{6}
\end{align*}
for sufficient large $n$.
So, with probability larger than $1-\frac{\alpha}{3}$, we have $$\left\|\frac{1}{n}\sum_{i=1}^n [\hat{\U}_i-\U_i][\hat{\U}_i-\U_i]^T\right\|_2\le\frac{1}{n}\sum_{i=1}^n  r_i^{-2}\left\|\hat{\bm \mu}-\bm \mu\right\|_2^2\le 2C_\alpha^2\zeta n^{-1}.$$

By Lemma \ref{le1}, we can rewrite
\begin{align*}
\hat{\U}_i-\U_i=r_i^{-1}[\I_d-\U_i\U_i^T]\hbth+\omega_i
\end{align*}
where $\|\omega_i\|_2\le C r_i^{-1-\delta}\|\hbth\|_2^{1+\delta}$. Thus,
\begin{align*}
\frac{2}{n}\sum_{i=1}^n [\hat{\U}_i-\U_i]\U_i^T =\frac{2}{n}\sum_{i=1}^n r_i^{-1}[\I_d-\U_i\U_i^T]\hbth\U_i^T+\frac{2}{n}\sum_{i=1}^n \omega_i\U_i^T.
\end{align*}
First, for any $\u\in \mathbb{S}^{d-1}$,
\begin{align*}
\frac{2}{n}\sum_{i=1}^n \u^T\omega_i\U_i^T\u&\le 2\sqrt{\frac{1}{n}\sum_{i=1}^n (\u^T\omega_i)^2\frac{1}{n}\sum_{i=1}^n (\u^T\U_i)^2}\\
&\le 2\sqrt{\frac{1}{n}\sum_{i=1}^n\|\omega_i\|_2^2}\le 2C \|\hbth\|_2^{1+\delta} \sqrt{\frac{1}{n}\sum_{i=1}^n r_i^{-2-2\delta}}.
\end{align*}
Similarly, by the Chebyshev inequality, we have
\begin{align*}
P\left(\frac{1}{n}\sum_{i=1}^n r_i^{-2-2\delta} \zeta_1^{-2-2\delta}\ge 2E(\nu_i^{2+2\delta})\right) \le \frac{\var(\nu_i^{2+2\delta})}{n[E(\nu_i^{2+2\delta})]^2} \le \frac{\alpha}{12}
\end{align*}
for sufficient large $n$. So, with probability larger than $1-\alpha/4$, we have
\begin{align*}
\left\|\frac{2}{n}\sum_{i=1}^n \omega_i\U_i^T\right\|_2&\le 2\sqrt{2}C \zeta_1^{1+\delta}\|\hbth\|_2^{1+\delta} [E(\nu_i^{2+2\delta})]^{1/2}\\
&\le 2\sqrt{2}C C_\alpha^{1+\delta} \zeta^{\frac{1+\delta}{4}}n^{-\frac{1}{2}(1+\delta)}.
\end{align*}
By Lemma \ref{mu}, we can write $\hbth=\zeta_1^{-1}\frac{1}{n}\sum_{i=1}^n \U_i+\zeta_1^{-1}\varrho$, where $\|\varrho\|_2=O_p(n^{-1})$. So
\begin{align*}
&\frac{2}{n}\sum_{i=1}^n r_i^{-1}[\I_d-\U_i\U_i^T]\hbth\U_i^T\\
=&\frac{2}{n^2}\sum_{i=1}^n \sum_{j=1}^n\zeta_1^{-1} r_i^{-1}[\I_d-\U_i\U_i^T] \U_j\U_i^T
+\frac{2}{n}\sum_{i=1}^n r_i^{-1}\zeta_1^{-1}[\I_d-\U_i\U_i^T]\varrho\U_i^T\\
=&\frac{2}{n^2}\sum_{i=1}^n \sum_{j\not=i}^n\nu_i[\I_d-\U_i\U_i^T] \U_j\U_i^T+\frac{2}{n}\sum_{i=1}^n \nu_i[\I_d-\U_i\U_i^T]\varrho\U_i^T\\
\doteq& B_1+B_2
\end{align*}
%
By the Chebyshev inequality, for any $\u\in \mathbb{S}^{d-1}$, we have
\begin{align*}
P\left(\frac{2}{n^2}\sum_{i=1}^n \sum_{j\not=i}^n\nu_i \u^T[\I_d-\U_i\U_i^T]\U_j\U_i^T\u \ge \sqrt{18/\alpha}n^{-1}\sigma_v \|\S\|_2\right) \le \frac{n^{-2}\sigma_v^2 (\u^T\S\u)^2}{18\alpha^{-1}n^{-2}\sigma_v^2\|\S\|_2^2} \le \frac{\alpha}{18}.
\end{align*}
So $P\left(\|B_1\|{ _2}\ge 3\sqrt{2}\alpha^{-1/2}n^{-1}\sigma_v \|\S\|_2\right)\le \frac{\alpha}{18}.$
Additionally,
\begin{align*}
&\left(\frac{1}{n}\sum_{i=1}^n \u^T\nu_i[\I_d-\U_i\U_i^T]\varrho\U_i^T\u\right)^2\\
&\le \left(\frac{1}{n}\sum_{i=1}^n(\u^T[\I_d-\U_i\U_i^T]\varrho)^2\right) \left(\frac{1}{n}\sum_{i=1}^n\nu_i^2(\U_i^T\u)^2\right)\\
&\le \frac{1}{n}\sum_{i=1}^n\nu_i^2  \|\rho\|_2^2=O_p(n^{-2}).
\end{align*}
So, for sufficient large $n$, there exist a constant $C_\rho$, such that $P\left(\|B_2\|_2 \ge C_\rho n^{-1}\right)\le \frac{\alpha}{18}$.
Finally, by the triangle inequality, for any $\alpha>0$, there exist a positive constant $C_S$, such that, for sufficient large $n$ and $\delta\in (0,1)$,
\begin{align*}
\|\hat{\S}-\S\|_2\le \|\mathbf{S}\|_2 \sqrt{\frac{4\left(1+r^*(\mathbf{S})\right)(\log d+\log (3    / \alpha))}{n}}+C_S n^{-\frac{1}{2}(1+\delta)}
\end{align*}
with probability larger than $1-\alpha$. \hfill$\Box$

\subsection{Proof of Corollary \ref{cor1}}
The Davis-Kahan inequality states that the approximation error of $\boldsymbol{u}_1(\widehat{\mathbf{S}})$ to $\boldsymbol{u}_1(\mathbf{S})$ is controlled by $\|\widehat{\mathbf{S}}-\mathbf{S}\|_2$ divided by the eigengap between $\lambda_1(\mathbf{S})$ and $\lambda_2(\mathbf{S})$ :
$$
\left|\sin \angle\left(\boldsymbol{u}_1(\widehat{\mathbf{S}}), \boldsymbol{u}_1(\mathbf{S})\right)\right| \leq \frac{2}{\lambda_1(\mathbf{S})-\lambda_2(\mathbf{S})}\|\widehat{\mathbf{S}}-\mathbf{S}\|_2.
$$
Thus, by Theorem \ref{th1}, we can directly obtain the result. \hfill$\Box$

\subsection{Proof of Theorem \ref{th2}}

\begin{lemma}\label{ub3} Suppose that $\boldsymbol{X} \sim E C_d(\boldsymbol{\mu}, \boldsymbol{\Sigma}, \xi)$ is elliptically distributed. For any $\boldsymbol{v} \in \mathbb{S}^{d-1}$, suppose that
\begin{align}\label{b3}
E \exp \left(t\left[\left(\boldsymbol{v}^T U(\boldsymbol{X})\right)^2-\boldsymbol{v}^T \mathbf{S} \boldsymbol{v}\right]\right) \leq \exp \left(\eta t^2\right), \quad \text { for } t \leq c_0 / \sqrt{\eta}
\end{align}
where $\eta>0$ only depends on the eigenvalues of $\boldsymbol{\Sigma}$ and $c_0$ is an absolute constant. We then have, with probability no smaller than $1-2 \alpha$, for large enough $n$,
$$
\sup _{\boldsymbol{v} \in \mathbb{S}^{d-1} \cap \mathbb{B}_0(s)}\left|\boldsymbol{v}^T(\widetilde{\mathbf{S}}-\mathbf{S}) \boldsymbol{v}\right| \leq 4\eta^{1 / 2} \sqrt{\frac{s(3+\log (d / s))+\log (1 / \alpha)}{n}}
$$
\end{lemma}
{\bf Proof:} Let $a \in \mathbb{Z}^{+}$be an integer no smaller than 1 and $J_a$ be any subset of $\{1, \ldots, d\}$ with cardinality $a$. For any $s$-dimensional sphere $\mathbb{S}^{s-1}$ equipped with Euclidean distance, we let $\mathcal{N}_\epsilon$ be a subset of $\mathbb{S}^{s-1}$ such that for any $\boldsymbol{v} \in \mathbb{S}^{s-1}$, there exists $\boldsymbol{u} \in \mathcal{N}_\epsilon$ subject to $\|\boldsymbol{u}-\boldsymbol{v}\|_2 \leq \epsilon$. It is known that the cardinal number of $\mathcal{N}_\epsilon$ has an upper bound: $\operatorname{card}\left(\mathcal{N}_\epsilon\right)<\left(1+\frac{2}{\epsilon}\right)^s$. Let $\mathcal{N}_{1 / 4}$ be a $(1 / 4)$-net of $\mathbb{S}^{s-1}$. We then have $\operatorname{card}\left(\mathcal{N}_{1 / 4}\right)$ is upper bounded by $9^s$. Moreover, for any symmetric matrix $\mathbf{M} \in \mathbb{R}^{s \times s}$, we have
$$
\sup _{\boldsymbol{v} \in \mathbb{S}^{s-1}}\left|\boldsymbol{v}^T \mathbf{M} \boldsymbol{v}\right| \leq \frac{1}{1-2 \epsilon} \sup _{\boldsymbol{v} \in \mathcal{N}_\epsilon}\left|\boldsymbol{v}^T \mathbf{M} \boldsymbol{v}\right| \text {, implying } \sup _{\boldsymbol{v} \in \mathbb{S}^{s-1}}\left|\boldsymbol{v}^T \mathbf{M} \boldsymbol{v}\right| \leq 2 \sup _{\boldsymbol{v} \in \mathcal{N}_{1 / 4}}\left|\boldsymbol{v}^T \mathbf{M} \boldsymbol{v}\right|
$$
Let $\beta>0$ be a quantity defined as $\beta:=(8 \eta)^{1 / 2} \sqrt{\frac{s(3+\log (d / s))+\log (1 / \alpha)}{n}}$. By the union bound, we have
$$
\begin{aligned}
& P\left(\sup _{\boldsymbol{b} \in S^{s-1}} \sup _{J_s \subset\{1, \cdots, d\}}\left|\boldsymbol{b}^T[\widetilde{\mathbf{S}}-\mathbf{S}]_{J_s, J_s} \boldsymbol{b}\right|>2 \beta\right) \leq P\left(\sup _{\boldsymbol{b} \in \mathcal{N}_{1 / 4}} \sup _{J_s \subset\{1, \cdots, d\}}\left|\boldsymbol{b}^T[\widetilde{\mathbf{S}}-\mathbf{S}]_{J_s, J_s} \boldsymbol{b}\right|>\beta\right) \\
\leq & 9^s\binom{d}{s} P\left(\left|\boldsymbol{b}^T[\widetilde{\mathbf{S}}-\mathbf{S}]_{J_s, J_s} \boldsymbol{b}\right|>(8 \eta)^{1 / 2} \sqrt{\frac{s(3+\log (d / s))+\log (1 / \alpha)}{n}}, \text { for any } \boldsymbol{b} \text { and } J_s\right)
\end{aligned}
$$
Thus, if we can show that for any $\boldsymbol{b} \in \mathbb{S}^{s-1}$ and $J_s$, we have
\begin{align}\label{c1}
P\left(\left|\boldsymbol{b}^T[\widetilde{\mathbf{S}}-\mathbf{S}]_{J_s, J_s} \boldsymbol{b}\right|>t\right) \leq 2 e^{-n t^2 /(4 \eta)}
\end{align}
for $\eta$ defined in Equation (\ref{b3}). Then, using the bound $\binom{d}{s}<(e d / s)^s$, we have
$$
\begin{aligned}
& 9^s\binom{d}{s} P\left(\left|\boldsymbol{b}^T[\widetilde{\mathbf{S}}-\mathbf{S}]_{J_s, J_s} \boldsymbol{b}\right|>(4 \eta)^{1 / 2} \sqrt{\frac{s(3+\log (d / s))+\log (1 / \alpha)}{n}}, \text { for any } \boldsymbol{b} \text { and } J\right) \le 2 \alpha.
\end{aligned}
$$
In fact, by the assumption (\ref{b3}) and Markov inequality, we have
\begin{align*}
&P\left(\boldsymbol{b}^T[\widetilde{\mathbf{S}}-\mathbf{S}]_{J_s, J_s} \boldsymbol{b}>t\right)\le E\left( e^{-nt^2/(2\eta)}\exp\left[tn/(2\eta)\boldsymbol{b}^T[\widetilde{\mathbf{S}}-\mathbf{S}]_{J_s, J_s} \boldsymbol{b}\right]\right)\\
=&e^{-nt^2/(2\eta)}E\left( \exp\left[tn/(2\eta)\boldsymbol{b}^T[\frac{1}{n}\sum_{i=1}^T\U_i\U_i^T-\mathbf{S}]_{J_s, J_s} \boldsymbol{b}\right]\right)\\
=&e^{-nt^2/(2\eta)}\left\{E(\exp((2\eta)^{-1}t\boldsymbol{b}^T[\U_i\U_i^T-\mathbf{S}]_{J_s, J_s}\boldsymbol{b}))\right\}^n\\
\le &e^{-nt^2/(2\eta)}\left\{E(\exp((2\eta)^{-1}t\boldsymbol{u}^T[\U_i\U_i^T-\mathbf{S}]\boldsymbol{u}))\right\}^n\\
\le &e^{-nt^2/(2\eta)}e^{(4\eta)^{-1} nt^2}\le e^{-nt^2/(4\eta)}
\end{align*}
for $t\le c_0\eta^{1/2}$. By symmetry, we can easily obtain the result (\ref{c1}).

{\bf Proof of Theorem \ref{th2}}: Note that $U(\X)$ has the same distribution as $S(X)=\frac{\X-\tilde \X}{\|\X-\tilde \X\|{ _2}}$ where $\X,\tilde \X \sim EC_d(\bmu,\bms,\xi)$ and are independent. By Lemma B.4 in \cite{han2018eca}, for any $\boldsymbol{v}=\left(v_1, \ldots, v_d\right)^T \in \mathbb{S}^{d-1}$, Equation (\ref{b3}) holds with
$$
\eta=\sup _{\boldsymbol{v} \in \mathbb{S}^{d-1}} 2\left\|\boldsymbol{v}^T U(\boldsymbol{X})\right\|_{\psi_2}^2+\|\mathbf{K}\|_2
$$
and
$$
\sup _{\boldsymbol{v} \in \mathbb{S}^{d-1}}\left\|\boldsymbol{v}^T U(\boldsymbol{X})\right\|_{\psi_2}=\sup _{\boldsymbol{v} \in \mathbb{S}^{d-1}}\left\|\frac{\sum_{i=1}^d v_i \lambda_i^{1 / 2}(\boldsymbol{\Sigma}) Y_i}{\sqrt{\sum_{i=1}^d \lambda_i(\boldsymbol{\Sigma}) Y_i^2}}\right\|_{\psi_2}
$$
where $\left(Y_1, \ldots, Y_d\right)^T \sim N_d\left(\mathbf{0}, \mathbf{I}_d\right)$ is standard Gaussian. Thus, by Lemma \ref{ub3}, when $(s \log (e d / s)+\log (1 / \alpha)) / n \rightarrow 0$, with probability at least $1-2 \alpha$, we have
$$
\begin{aligned}
\|\widetilde{\mathbf{S}}-\mathbf{S}\|_{2, s} \leq &  C_0 \left(\sup _{\boldsymbol{v} \in \mathbb{S}^{d-1}} 2\left\|\boldsymbol{v}^T U(\boldsymbol{X})\right\|_{\psi_2}^2+\|\mathbf{S}\|_2\right)^{1/2}\sqrt{\frac{s(3+\log (d / s))+\log (1 / \alpha)}{n}}.
\end{aligned}
$$
Taking the same procedure as Lemma \ref{mu}, we can have $\|\hbth\|_{2,s}=O_p(\sqrt{\frac{s}{d}}\zeta_1^{-1}n^{-1/2})$. So taking the same procedure as the proof of Theorem \ref{th1}, we have, there exist some positive constant $C_1>0$,
\begin{align*}
\|\widehat{\S}-\widetilde{\S}\|_{2,s}\le C_1 \left(\frac{nd}{s}\right)^{-\frac{1}{2}(1+\delta)}
\end{align*}
for sufficient large $n$ and with probability larger than $1-\alpha$. Then, by the triangle inequality, we obtain the first result. Specially, if $\operatorname{rank}(\boldsymbol{\Sigma})=q$ and $\left\|\boldsymbol{u}_1(\boldsymbol{\Sigma})\right\|_0 \leq s$, by Theorem 4.2 in \cite{han2018eca}, we have
$$
\sup _{\boldsymbol{v} \in \mathbb{S}^{d-1}}\left\|\boldsymbol{v}^T U(\boldsymbol{X})\right\|_{\psi_2} \leq \sqrt{\frac{\lambda_1(\boldsymbol{\Sigma})}{\lambda_q(\boldsymbol{\Sigma})} \cdot \frac{2}{q}} \wedge 1.
$$
So we can easily obtain the second result. \hfill$\Box$

\subsection{Proof of Corollary \ref{cor2}}
The Davis-Kahan inequality and Theorem \ref{th2}, we can directly obtain the result. \hfill$\Box$

\subsection{Proof of Theorem \ref{thad}}
Using the result in Theorem \ref{th2}, we have for any $\boldsymbol{v} \in \mathbb{S}^{d-1}$,
\begin{align}\label{sb}
\left\|\boldsymbol{v}^T U(\boldsymbol{X})\right\|_{\psi_2} \leq \sqrt{\frac{\lambda_1(\boldsymbol{\Sigma})}{\lambda_q(\boldsymbol{\Sigma})} \cdot \frac{2}{q}}.
\end{align}
So, for any $j,k\in \{1,\cdots,d\}$, we have
\begin{align*}
\|\bm e_j^T U(\X_i)U(\X_i)^T \bm e_k\|_{\psi_1}\leq {\frac{\lambda_1(\boldsymbol{\Sigma})}{\lambda_q(\boldsymbol{\Sigma})} \cdot \frac{8}{q}}
\end{align*}
where $\bm e_j=(0,\cdots,0,1,0,\cdots,0)$ with the $j$-th element being one and the others are zeros. So by the Bernstein inequality, we have
\begin{align*}
P\left(|\bm e_j^T U(\X_i)U(\X_i)^T \bm e_k-\S_{jk}|\ge t\right)\le \exp\left(-\frac{nt^2}{8 C\left(8\left(\lambda_1(\boldsymbol{\Sigma}) / q \lambda_q(\boldsymbol{\Sigma})\right)+\mathbf{S}_{j k}\right)^2}\right)
\end{align*}
for $t<2 C c\left(8 \lambda_1(\boldsymbol{\Sigma}) / q \lambda_q(\boldsymbol{\Sigma})+\mathbf{K}_{j k}\right)$.
So,
\begin{align*}
P\left(\|\tilde \S-\S\|_{max}\ge t\right)\le d^2 \exp\left(-\frac{nt^2}{8 C\left(8 \lambda_1(\boldsymbol{\Sigma}) / q \lambda_q(\boldsymbol{\Sigma})+\|\mathbf{S}\|_{\max }\right)^2}\right).
\end{align*}
Then, with probability larger than $1-\alpha^2$, for sufficient large $n$, we have
\begin{align*}
\|\tilde \S-\S\|_{max} \le  4 \sqrt{C}\left(\frac{8 \lambda_1(\boldsymbol{\Sigma})}{q \lambda_q(\boldsymbol{\Sigma})}+\|\mathbf{S}\|_{\max }\right)\sqrt{\frac{\log d+\log (1/\alpha)}{n}}.
\end{align*}

Next, we will show the bound of $\|\hat{\S}-\tilde{\S}\|_{max}$. Because
\begin{align*}
\hat{\S}-\tilde{\S}=&\frac{1}{n}\sum_{i=1}^n [\hat{\U}_i\hat{\U}_i^T-\U_i\U_i^T]\\
=&\frac{2}{n}\sum_{i=1}^n [\hat{\U}_i-\U_i]\U_i^T+\frac{1}{n}\sum_{i=1}^n [\hat{\U}_i-\U_i][\hat{\U}_i-\U_i]^T\doteq J_1+J_2
\end{align*}
Because
\begin{align*}
\hat{\U}_i-\U_i=&\frac{\X_i-\hat\bmu}{\|\X_i-\hat\bmu\|{ _2}}-\frac{\X_i-\bmu}{\|\X_i-\bmu\|{ _2}}\\
=&\frac{\X_i-\hat\bmu}{\|\X_i-\hat\bmu\|{ _2}}-\frac{\X_i-\hat\bmu}{\|\X_i-\bmu\|{ _2}}+\frac{\X_i-\hat\bmu}{\|\X_i-\bmu\|{ _2}}-\frac{\X_i-\bmu}{\|\X_i-\bmu\|{ _2}}\\
=&(\X_i-\hat\bmu)(\hat r_i^{-1}-{r}_i^{-1})-(\hat\bmu-\bmu)r_i^{-1}\\
=&\U_i(r_i\hat r_i^{-1}-1)-(\hat{\bmu}-\bmu)(\hat r_i^{-1}-{r}_i^{-1})-(\hat\bmu-\bmu)r_i^{-1}.
\end{align*}
So
\begin{align*}
\left\|\frac{J_1}{2}\right\|_{max}&\le \left\|\frac{1}{n}\sum_{i=1}^n(r_i\hat r_i^{-1}-1)\U_i\U_i^T\right\|_{max}+\left\|\frac{1}{n}\sum_{i=1}^n(\hat r_i^{-1}-{r}_i^{-1})(\hat{\bmu}-\bmu)\U_i^T\right\|_{max}\\
&+\left\|\frac{1}{n}\sum_{i=1}^n(\hat{\bmu}-\bmu)r_i^{-1}\U_i^T\right\|_{max}\\
&\doteq J_{11}+J_{12}+J_{13}.
\end{align*}
First,
\begin{align*}
J_{11}\le \max_{1\le i\le T}|r_i\hat r_i^{-1}-1|\|\tilde{\S}\|_{max}.
\end{align*}
Because $|\hat r_i-r_i|\le \|\hat{\bmu}-\bmu\|{ _2}$, so $|r_i\hat{r}_i^{-1}-1|\le \frac{\|\hat{\bmu}-\bmu\|{ _2}}{r_i-\|\hat{\bmu}-\bmu\|{ _2}}$. Then,
\begin{align*}
\max_{1\le i\le T}|r_i\hat r_i^{-1}-1| \le \zeta_1\|\hat{\bmu}-\bmu\|{ _2} \max_{1\le i\le T} (\zeta_1r_i-\zeta_1\|\hat{\bmu}-\bmu\|{ _2})^{-1}.
\end{align*}
By Lemma \ref{mu}, we have $\zeta_1\|\hat{\bmu}-\bmu\|{ _2}=O_p(n^{-1/2})$. And by the sub-gaussian assumption of $\nu_i$, we have $\max_{1\le i\le n}\nu_i=O_p(\log^{1/2} n)$. So $\max_{1\le i\le T}|r_i\hat r_i^{-1}-1|=O_p(\sqrt{\frac{\log n}{n}})$. In addition $\|\tilde \S\|_{max}\le \|\tilde{\S}-\S\|_{max}+\|\S\|_{max}$. So, with probability larger than $1-\alpha$,
\begin{align*}
J_{11}\le C\sqrt{\frac{\log n}{n}}\left(\|\S\|_{max}+ 4 \sqrt{C}\left(\frac{8 \lambda_1(\boldsymbol{\Sigma})}{q \lambda_q(\boldsymbol{\Sigma})}+\|\S\|_{max }\right)\sqrt{\frac{\log d+\log (1/\alpha)}{n}}\right)
\end{align*}
for sufficient large $C$ and $n$.

Obviously,
\begin{align*}
J_{13}\le \zeta_1\|\hat\bmu-\bmu\|_{\infty}\left\|\frac{1}{n}\sum_{i=1}^T \nu_i\U_i\right\|_{\infty}.
\end{align*}
By (\ref{sb}) and $\nu_i$ is also sub-gaussian variable with parameter $K_\nu$, we have
\begin{align*}
\|\nu_i\U_i^T \v\|_{\psi_1}\le \|\nu_i\|_{\psi_2}\|\U_i^T\v\|_{\psi_2}\le K_\nu\sqrt{\frac{\lambda_1(\boldsymbol{\Sigma})}{\lambda_q(\boldsymbol{\Sigma})} \cdot \frac{2}{q}}.
\end{align*}
So, by the Bernstein inequality, we have
\begin{align*}
P\left(\frac{1}{n}\sum_{i=1}^T\nu_i\U_i^T e_k\ge t\right)\le \exp\left(-\frac{nt^2}{8CK_\nu^2\lambda_1(\boldsymbol{\Sigma}) / q \lambda_q(\boldsymbol{\Sigma})}\right)
\end{align*}
for $t\le c K_\nu \sqrt{\lambda_1(\boldsymbol{\Sigma}) / q \lambda_q(\boldsymbol{\Sigma})}$ for some positive constant $c,C$. Then,
\begin{align*}
P\left(\left\|\frac{1}{n}\sum_{i=1}^T \nu_i\U_i\right\|_{\infty}\ge t\right)\le d\exp\left(-\frac{nt^2}{8CK_\nu^2\lambda_1(\boldsymbol{\Sigma}) / q \lambda_q(\boldsymbol{\Sigma})}\right) \le \alpha
\end{align*}
by setting $t=\sqrt{\frac{8CK_\nu^2\lambda_1(\boldsymbol{\Sigma})}{q \lambda_q(\boldsymbol{\Sigma})}}\sqrt{\frac{\log d+\log (1/\alpha)}{n}}$.
Similar to the above arguments, we have
\begin{align*}
P\left(\left\|\frac{1}{n}\sum_{i=1}^T \U_i\right\|_{\infty}\ge \sqrt{\frac{8C\lambda_1(\boldsymbol{\Sigma})}{q \lambda_q(\boldsymbol{\Sigma})}}\sqrt{\frac{\log d+\log (1/\alpha)}{n}}\right)\le \alpha.
\end{align*}
By the equation $\sum_{i=1}^{n}\hat\U_i=0$, we have
\begin{align*}
\zeta_1(\hat\bmu-\bmu)=&\frac{1}{n}\sum_{i=1}^n \U_i+\frac{1}{n}\sum_{i=1}^n (r_i\hat{r}_i^{-1}-1) \U_i +\left(\frac{\zeta_1}{\frac{1}{n}\sum_{i=1}^n \hat{r}_i^{-1}}-1\right)\frac{1}{n}\sum_{i=1}^n \U_i\\
&+\left(\frac{\zeta_1}{\frac{1}{n}\sum_{i=1}^n \hat{r}_i^{-1}}-1\right)\frac{1}{n}\sum_{i=1}^n \U_i(r_i\hat{r}_i^{-1}-1)\\
\doteq & B_1+B_2+B_3+B_4.
\end{align*}
By the sub-gaussian assumption of $\nu_i$, we have $\|B_i\|_{\infty}=O_p\left(\sqrt{\frac{\log n}{n}}\right)\|B_1\|_\infty$, $i=2,3,4$.
So
\begin{align*}
\zeta_1\|\hat\bmu-\bmu\|_{\infty}\le \sqrt{\frac{9C\lambda_1(\boldsymbol{\Sigma})}{q \lambda_q(\boldsymbol{\Sigma})}}\sqrt{\frac{\log d+\log (1/\alpha)}{n}}
\end{align*}
with probability larger than $1-\alpha$ for sufficient large $n$ and $C$. Then,
\begin{align*}
J_{13} \le {\frac{C\lambda_1(\boldsymbol{\Sigma})}{q \lambda_q(\boldsymbol{\Sigma})}}{\frac{\log d+\log (1/\alpha)}{n}}
\end{align*}
with probability larger than $1-2\alpha$ for sufficient large $n$ and $C$.
Finally,
\begin{align*}
J_{12}\le \max_{1\le i\le T}|r_i\hat r_i^{-1}-1| J_{13}
\end{align*}
which implies $J_{12}$ is a smaller order than $J_{13}$. So, by the triangle inequality, we obtain that
$\|J_1\|_{max}=o_p(\|\tilde{\S}-\S\|_{max})$. Similarly, we also can prove that $\|J_2\|_{max}=o_p(\|\tilde{\S}-\S\|_{max})$. So, we have
\begin{align*}
\|\hat \S-\S\|_{max} \le  C\left(\frac{8 \lambda_1(\boldsymbol{\Sigma})}{q \lambda_q(\boldsymbol{\Sigma})}+\|\mathbf{S}\|_{\max }\right)\sqrt{\frac{\log d+\log (1/\alpha)}{n}}.
\end{align*}
with probability larger than $1-\alpha^2$. The rest proves are all similar to the proof of Theorem 5.3, Corollary 5.3 and Theorem 5.4 in \cite{han2018eca}. So we omit the details here. \hfill$\Box$

\bibliographystyle{apa}

\bibliography{Referspca}

\begin{thebibliography}{}

\bibitem[\protect\astroncite{Arcones}{1998}]{arcones1998asymptotic}
Arcones, M. (1998).
\newblock Asymptotic theory for m-estimators over a convex kernel.
\newblock {\em Econometric Theory}, 14:387--422.

\bibitem[\protect\astroncite{Bai et~al.}{1990}]{bai1990asymptotic}
Bai, Z., Chen, R., Miao, B., and Rao, C. (1990).
\newblock Asymptotic theory of least distances estimate in multivariate linear
  models.
\newblock {\em Statistics}, 4:503--519.

\bibitem[\protect\astroncite{Baik and Silverstein}{2006}]{Baik2006}
Baik, J. and Silverstein, J.~W. (2006).
\newblock Eigenvalues of large sample covariance matrices of spiked population
  models.
\newblock {\em Journal of Multivariate Analysis}, 97:1382--1408.

\bibitem[\protect\astroncite{Balcan et~al.}{2016}]{balcan2016communication}
Balcan, M.~F., Liang, Y., Song, L., Woodruff, D., and Xie, B. (2016).
\newblock Communication efficient distributed kernel principal component
  analysis.
\newblock In {\em Proceedings of the 22nd ACM SIGKDD International Conference
  on Knowledge Discovery and Data Mining}, pages 725--734.

\bibitem[\protect\astroncite{Bickel and Levina}{2008}]{Bickel2008}
Bickel, P. and Levina, E. (2008).
\newblock Regularized estimation of large covariance matrices.
\newblock {\em The Annals of Statistics}, 36:199--227.

\bibitem[\protect\astroncite{Birnbaum et~al.}{2013}]{Birnbaum2013}
Birnbaum, A., Johnstone, I.~M., Nadler, B., and Paul, D. (2013).
\newblock Minimax bounds for sparse pca with noisy high-dimensional data.
\newblock {\em The Annals of statistics}, 41:1055--1084.

\bibitem[\protect\astroncite{Cai et~al.}{2013}]{Cai2013}
Cai, T.~T., Ma, Z., and Wu, Y. (2013).
\newblock Sparse pca: Optimal rates and adaptive estimation.
\newblock {\em The Annals of Statistics}, 41:3074--3110.

\bibitem[\protect\astroncite{Chan et~al.}{2015}]{chan2015pcanet}
Chan, T.-H., Jia, K., Gao, S., Lu, J., Zeng, Z., and Ma, Y. (2015).
\newblock Pcanet: A simple deep learning baseline for image classification?
\newblock {\em IEEE transactions on image processing}, 24(12):5017--5032.

\bibitem[\protect\astroncite{Chen et~al.}{2022}]{chen2022high}
Chen, X., Zhang, J., and Zhou, W. (2022).
\newblock High-dimensional elliptical sliced inverse regression in non-gaussian
  distributions.
\newblock {\em Journal of Business \& Economic Statistics}, 40(3):1204--1215.

\bibitem[\protect\astroncite{Cheng et~al.}{2019}]{cheng2019testing}
Cheng, G., Liu, B., Peng, L., Zhang, B., and Zheng, S. (2019).
\newblock Testing the equality of two high-dimensional spatial sign covariance
  matrices.
\newblock {\em Scandinavian Journal of Statistics}, 46(1):257--271.

\bibitem[\protect\astroncite{Croux et~al.}{2013}]{Croux2013}
Croux, C., Filzmoser, P., and Fritz, H. (2013).
\newblock Robust sparse principal component analysis.
\newblock {\em Technometrics}, 55:202--214.

\bibitem[\protect\astroncite{Croux et~al.}{2002}]{Croux2002}
Croux, C., Ollila, E., and Oja, H. (2002).
\newblock Sign and rank covariance matrices: Statistical properties and
  application to principal components analysis.
\newblock {\em Statistics for Industry and Technology}, pages 257--269.

\bibitem[\protect\astroncite{Davis and Kahan}{1970}]{Davis1970}
Davis, C. and Kahan, W.~M. (1970).
\newblock The rotation of eigenvectors by a perturbation. iii.
\newblock {\em SIAM Journal on Numerical Analysis}, 7:1--46.

\bibitem[\protect\astroncite{Fan et~al.}{2008}]{Fan2008}
Fan, J., Fan, Y., and Lv, J. (2008).
\newblock High dimensional covariance matrix estimation using a factor model.
\newblock {\em Journal of Econometrics}, 147:186--197.

\bibitem[\protect\astroncite{Feng}{2018}]{feng2018power}
Feng, L. (2018).
\newblock Power comparison between high dimensional t-test, sign, and signed
  rank tests.
\newblock {\em arXiv preprint arXiv:1812.10625}.

\bibitem[\protect\astroncite{Feng and Liu}{2017}]{fl2017}
Feng, L. and Liu, B. (2017).
\newblock High-dimensional rank tests for sphericity.
\newblock {\em Journal of Multivariate Analysis}, 155:217--233.

\bibitem[\protect\astroncite{Feng et~al.}{2021}]{f2021}
Feng, L., Liu, B., and Ma, Y. (2021).
\newblock An inverse norm sign test of location parameter for high-dimensional
  data.
\newblock {\em Journal of Business \& Economic Statistics}, 39(3):807--815.

\bibitem[\protect\astroncite{Feng and Sun}{2016}]{feng2016}
Feng, L. and Sun, F. (2016).
\newblock Spatial-sign based high-dimensional location test.
\newblock {\em Electronic Journal of Statistics}, 10:2420--2434.

\bibitem[\protect\astroncite{Feng et~al.}{2016}]{Feng2016Multivariate}
Feng, L., Zou, C., and Wang, Z. (2016).
\newblock Multivariate-sign-based high-dimensional tests for the two-sample
  location problem.
\newblock {\em Journal of the American Statistical Association},
  111(514):721--735.

\bibitem[\protect\astroncite{Han and Liu}{2014}]{Han2014}
Han, F. and Liu, H. (2014).
\newblock Scale-invariant sparse pca on high-dimensional meta-elliptical data.
\newblock {\em Journal of the American Statistical Association}, 109:275--287.

\bibitem[\protect\astroncite{Han and Liu}{2018}]{han2018eca}
Han, F. and Liu, H. (2018).
\newblock Eca: High-dimensional elliptical component analysis in non-gaussian
  distributions.
\newblock {\em Journal of the American Statistical Association},
  113(521):252--268.

\bibitem[\protect\astroncite{He et~al.}{2022}]{he2022large}
He, Y., Kong, X., Yu, L., and Zhang, X. (2022).
\newblock Large-dimensional factor analysis without moment constraints.
\newblock {\em Journal of Business \& Economic Statistics}, 40(1):302--312.

\bibitem[\protect\astroncite{Huang et~al.}{2023}]{h2022}
Huang, X., Liu, B., Zhou, Q., and Feng, L. (2023).
\newblock A high-dimensional inverse norm sign test for two-sample location
  problems.
\newblock {\em Canadian Journal of Statistics}, 51(4):1004--1033.

\bibitem[\protect\astroncite{Hubert et~al.}{2016}]{hubert2016sparse}
Hubert, M., Reynkens, T., Schmitt, E., and Verdonck, T. (2016).
\newblock Sparse pca for high-dimensional data with outliers.
\newblock {\em Technometrics}, 58(4):424--434.

\bibitem[\protect\astroncite{Hubert et~al.}{2005}]{hubert2005robpca}
Hubert, M., Rousseeuw, P.~J., and Vanden~Branden, K. (2005).
\newblock Robpca: a new approach to robust principal component analysis.
\newblock {\em Technometrics}, 47(1):64--79.

\bibitem[\protect\astroncite{Johnstone and Lu}{2009}]{Johnstone2009}
Johnstone, I.~M. and Lu, A.~Y. (2009).
\newblock On consistency and sparsity for principal components analysis in high
  dimensions.
\newblock {\em Journal of the American Statistical Association}, 104:682--693.

\bibitem[\protect\astroncite{Jolliffe et~al.}{2003}]{Jolliffe2003modified}
Jolliffe, I., Trendafilov, N., and Uddin, M. (2003).
\newblock A modified principal component technique based on the lasso.
\newblock {\em Journal of Computational and Graphical Statistics}, 12:531--547.

\bibitem[\protect\astroncite{Journee et~al.}{2010}]{journee2010generalized}
Journee, M., Nesterov, Y., Richtarik, P., and Sepulchre, R. (2010).
\newblock Generalized power method for sparse principal component analysis.
\newblock {\em Journal of Machine Learning Research}, 11:517--553.

\bibitem[\protect\astroncite{Lan and Du}{2019}]{lan2019a}
Lan, W. and Du, L. (2019).
\newblock A factor-adjusted multiple testing procedure with application to
  mutual fund selection.
\newblock {\em Journal of Business \& Economic Statistics}, 37(1):147--157.

\bibitem[\protect\astroncite{LeCun et~al.}{2002}]{lecun2002gradient}
LeCun, Y., Bottou, L., Bengio, Y., and Haffner, P. (2002).
\newblock Gradient-based learning applied to document recognition.
\newblock {\em Proceedings of the IEEE}, 86(11):2278--2324.

\bibitem[\protect\astroncite{Li and Xu}{2022}]{li2022asymptotic}
Li, W. and Xu, Y. (2022).
\newblock Asymptotic properties of high-dimensional spatial median in
  elliptical distributions with application.
\newblock {\em Journal of Multivariate Analysis}, 190:104975.

\bibitem[\protect\astroncite{Liu et~al.}{2023}]{liu2023high}
Liu, B., Feng, L., and Ma, Y. (2023).
\newblock High-dimensional alpha test of linear factor pricing models with
  heavy-tailed distributions.
\newblock {\em Statistica Sinica}, 33:1389--1410.

\bibitem[\protect\astroncite{Locantore et~al.}{1999}]{locantore1999robust}
Locantore, N., Marron, J., Simpson, D., Tripoli, N., Zhang, J., Cohen, K.,
  Boente, G., Fraiman, R., Brumback, B., Croux, C., et~al. (1999).
\newblock Robust principal component analysis for functional data.
\newblock {\em Test}, 8:1--73.

\bibitem[\protect\astroncite{Lounici}{2014}]{lounici2014high}
Lounici, K. (2014).
\newblock High-dimensional covariance matrix estimation with missing
  observations.
\newblock {\em Bernoulli}, 20:1029--1058.

\bibitem[\protect\astroncite{Ma}{2013}]{ma2013sparse}
Ma, Z. (2013).
\newblock Sparse principal component analysis and iterative thresholding.
\newblock {\em The Annals of Statistics}, 41:772--801.

\bibitem[\protect\astroncite{Mackey}{2008}]{mackey2008}
Mackey, L. (2008).
\newblock Deflation methods for sparse pca.
\newblock In {\em Advances in Neural Information Processing Systems},
  volume~21, pages 1017--1024.

\bibitem[\protect\astroncite{Marden}{1999}]{marden1999some}
Marden, J. (1999).
\newblock Some robust estimates of principal components.
\newblock {\em Statistics and Probability Letters}, 43:349--359.

\bibitem[\protect\astroncite{Nadler}{2008}]{Nadler2008}
Nadler, B. (2008).
\newblock Finite sample approximation results for principal component analysis:
  A matrix perturbation approach.
\newblock {\em The Annals of Statistics}, 36:2791--2817.

\bibitem[\protect\astroncite{Neter et~al.}{1996}]{neter1996}
Neter, J., Kutner, M., Wasserman, W., and Nachtsheim, C. (1996).
\newblock {\em Applied Linear Statistical Models}, volume~4.
\newblock Irwin, Chicago, IL.

\bibitem[\protect\astroncite{Oja}{2010}]{Oja2010Multivariate}
Oja, H. (2010).
\newblock {\em Multivariate nonparametric methods with R: an approach based on
  spatial signs and ranks}.
\newblock Springer Science \& Business Media.

\bibitem[\protect\astroncite{Paindaveine and
  Verdebout}{2016}]{paindaveine2016high}
Paindaveine, D. and Verdebout, T. (2016).
\newblock On high-dimensional sign tests.
\newblock {\em Bernoulli}, 22(3):1745--1769.

\bibitem[\protect\astroncite{Paul}{2007}]{Paul2007}
Paul, D. (2007).
\newblock Asymptotics of sample eigenstructure for a large dimensional spiked
  covariance model.
\newblock {\em Statistica Sinica}, pages 1617--1642.

\bibitem[\protect\astroncite{Shen and Huang}{2008}]{Shen2008}
Shen, H. and Huang, J.~Z. (2008).
\newblock Sparse principal component analysis via regularized low rank matrix
  approximation.
\newblock {\em Journal of multivariate analysis}, 99:1015--1034.

\bibitem[\protect\astroncite{Taskinen et~al.}{2012}]{Taskinen2012}
Taskinen, S., Koch, I., and Oja, H. (2012).
\newblock Robustifying principal component analysis with spatial sign vectors.
\newblock {\em Statistics and Probability Letters}, 82:765--774.

\bibitem[\protect\astroncite{Tropp}{2012}]{tropp2012}
Tropp, J.~A. (2012).
\newblock User-friendly tail bounds for sums of random matrices.
\newblock {\em Foundations of Computational Mathematics}, 12:389--434.

\bibitem[\protect\astroncite{Vershynin}{2010}]{vershynin2010}
Vershynin, R. (2010).
\newblock {\em Introduction to the Non-Asymptotic Analysis of Random Matrices},
  pages 210--268.
\newblock Cambridge University Press.

\bibitem[\protect\astroncite{Visuri et~al.}{2000}]{visuri2000sign}
Visuri, S., Koivunen, V., and Oja, H. (2000).
\newblock Sign and rank covariance matrices.
\newblock {\em Journal of Statistical Planning and Inference}, 91(2):557--575.

\bibitem[\protect\astroncite{Vu and Lei}{2012}]{VuLei2012}
Vu, V. and Lei, J. (2012).
\newblock Minimax rates of estimation for sparse pca in high dimensions.
\newblock In {\em Proceedings of the 15th International Conference on
  Artificial Intelligence and Statistics (AISTATS)}, volume~22, pages
  1278--1286.

\bibitem[\protect\astroncite{Vu et~al.}{2013}]{VuChoLeiRohe2013}
Vu, V.~Q., Cho, J., Lei, J., and Rohe, K. (2013).
\newblock Fantope projection and selection: A near-optimal convex relaxation of
  sparse pca.
\newblock In {\em Advances in Neural Information Processing Systems},
  volume~26, pages 2670--2678.

\bibitem[\protect\astroncite{Vu and Lei}{2013}]{VuLei2013}
Vu, V.~Q. and Lei, J. (2013).
\newblock Minimax sparse principal subspace estimation in high dimensions.
\newblock {\em The Annals of Statistics}, 41:2905--2947.

\bibitem[\protect\astroncite{Wang et~al.}{2015}]{wang2015high}
Wang, L., Peng, B., and Li, R. (2015).
\newblock A high-dimensional nonparametric multivariate test for mean vector.
\newblock {\em Journal of the American Statistical Association},
  110(512):1658--1669.

\bibitem[\protect\astroncite{Wang et~al.}{2013}]{WangHanLiu2013}
Wang, Z., Han, F., and Liu, H. (2013).
\newblock Sparse principal component analysis for high dimensional vector
  autoregressive models.
\newblock {\em arXiv:1307.0164}.

\bibitem[\protect\astroncite{Wedin}{1972}]{Wedin1972}
Wedin, P.-A. (1972).
\newblock Perturbation bounds in connection with singular value decomposition.
\newblock {\em BIT Numerical Mathematics}, 12:99--111.

\bibitem[\protect\astroncite{Witten et~al.}{2009}]{Witten2009}
Witten, D.~M., Tibshirani, R., and Hastie, T. (2009).
\newblock A penalized matrix decomposition, with applications to sparse
  principal components and canonical correlation analysis.
\newblock {\em Biostatistics}, 10:515--534.

\bibitem[\protect\astroncite{Yang and Du}{2025}]{yang2025}
Yang, X. and Du, L. (2025).
\newblock Multiple testing under high-dimensional dynamic factor model.

\bibitem[\protect\astroncite{Yuan and Zhang}{2013}]{YuanZhang2013}
Yuan, X. and Zhang, T. (2013).
\newblock Truncated power method for sparse eigenvalue problems.
\newblock {\em Journal of Machine Learning Research}, 14:899--925.

\bibitem[\protect\astroncite{Zhang et~al.}{2022}]{zhang2022robust}
Zhang, X., Zhao, P., and Feng, L. (2022).
\newblock Robust sphericity test in the panel data model.
\newblock {\em Statistics \& Probability Letters}, 182:109304.

\bibitem[\protect\astroncite{Zhang and El~Ghaoui}{2011}]{ZhangElGhaoui2011}
Zhang, Y. and El~Ghaoui, L. (2011).
\newblock Large-scale sparse principal component analysis with application to
  text data.
\newblock In {\em Advances in Neural Information Processing Systems},
  volume~24, pages 532--539.

\bibitem[\protect\astroncite{Zhao et~al.}{2023}]{zhao2023spatial}
Zhao, P., Chen, D., and Wang, Z. (2023).
\newblock Spatial-sign based high dimensional white noises test.
\newblock {\em arXiv preprint arXiv:2303.10641}.

\bibitem[\protect\astroncite{Zou et~al.}{2014}]{zou2014multivariate}
Zou, C., Peng, L., Feng, L., and Wang, Z. (2014).
\newblock Multivariate sign-based high-dimensional tests for sphericity.
\newblock {\em Biometrika}, 101(1):229--236.

\bibitem[\protect\astroncite{Zou et~al.}{2006}]{Zou2006}
Zou, H., Hastie, T., and Tibshirani, R. (2006).
\newblock Sparse principal component analysis.
\newblock {\em Journal of computational and graphical statistics}, 15:265--286.

\bibitem[\protect\astroncite{Zou and Xue}{2018}]{Zou2018}
Zou, H. and Xue, L. (2018).
\newblock A selective overview of sparse principal component analysis.
\newblock {\em Proceedings of the IEEE}, 106:1311--1320.

\end{thebibliography}

\end{document}